\documentclass{article}
\usepackage{amssymb}
\usepackage{amsmath}
\usepackage{palatino}
\newtheorem{Definition}{Definition}

\begin{document}

%\begin{titlepage}

\title{Coisotropic deformations of associative algebras
and dispersionless integrable hierarchies}

\author{B.G. Konopelchenko$^{1}$ and F. Magri$^{2}$
\\
$^{1}$\small{Dipartimento di Fisica, Universita' di Lecce and
Sezione
INFN,}\\  \small{73100 Lecce, Italy}\\
$^{2}$\small{Dipartimento di Matematica ed Applicazioni,
Universita' di Milano Bicocca,}\\ \small{20126 Milano, Italy}}

\date{}
\maketitle

\begin{abstract}
The paper is an inquiry of the algebraic foundations of the theory
of dispersionless integrable hierarchies, like the dispersionless
KP and modified KP hierarchies and the universal Whitham's
hierarchy of genus zero. It stands out for the idea of
interpreting these hierarchies as equations of coisotropic
deformations for the structure constants of certain associative
algebras. It discusses the link between the structure constants
and the Hirota's tau function, and shows that the dispersionless
Hirota's bilinear equations are, within this approach, a way of
writing the associativity conditions for the structure constants
in terms of the tau function. It also suggests a simple
interpretation of the algebro-geometric construction of the
universal Whitham's equations of genus zero due to Krichever.
\end{abstract}

%\end{titlepage}

\newpage

\section{Introduction.}

The purpose of this paper is to introduce
the concept of "coisotropic deformations" of associative
algebras and to show its relevance for the theory
of integrable hierarchies of dispersionless PDE's.

The concept of coisotropic deformation originates from
a melting of ideas borrowed from commutative algebra and
Hamiltonian mechanics. From commutative algebra comes
the idea of structure constants $C_{kj}^{l}$ defining the
table of multiplication
\begin{equation}
p_{j}p_{k} = \sum_{l=1}^{n}C_{jk}^{l}p_{l}
\end{equation}
of a commutative associative algebra with unity.
They obey the commutativity conditions
\begin{equation}\label{commutativity equations}
C_{jk}^{l}=C_{kj}^{l}
\end{equation}
and the associativity conditions
\begin{equation}\label{associativity equations}
\sum_{l=1}^{n}C_{jk}^{l}C_{lm}^{p}=\sum_{l=1}^{n}C_{mk}^{l}C_{lj}^{p} .
\end{equation}
Furthermore, if the algebra is infinite-dimensional, they
are assumed to vanish for l sufficiently large,
so that the sums are always over a finite number of terms.
From deformation theory comes the idea of regarding
the structure constants $C_{kj}^{l}$ as functions
of a certain number of deformation parameters $x_{j}$. From Hamiltonian
mechanics come the ideas of introducing a deformation
parameter $x_{j}$ for each generator $p_{j}$ of the algebra,
and of considering the pairs $(x_{j},p_{j})$ as pairs of conjugate
canonical variables.

In this frame, the characteristic trait
of the theory of coisotropic deformations is to associate
with the structure constants $C_{jk}^{l}(x_{1},x_{2},\ldots,x_{n})$
the set of quadratic Hamiltonians
\begin{equation}\label{eq: hamiltonians}
f_{jk}=-p_{j}p_{k}+\sum_{l=1}^{n}C_{jk}^{l}(x_{1},x_{2},\ldots,x_{n})p_{l} ,
\end{equation}
the polynomial ideal
\begin{equation}\label{eq: ideal}
J=<f_{jk}>
\end{equation}
generated by these Hamiltonians, and the submanifold
\begin{equation}
\Gamma=\{ (x_{j},p_{j})\in{R^{2n}}\quad | f_{jk}=0 \}
\end{equation}
where these Hamiltonians vanish. It lives in $R^{2n}$ endowed with
the canonical Poisson bracket

\begin{equation*}
\{ f,g \} =\sum_{i=1}^{i=n} \left(\frac{\partial f}{\partial
x_{i}}\frac{\partial g }{\partial p_{i} }- \frac{\partial g
}{\partial x_{i} }\frac{\partial f }{\partial p_{i} }\right).
\end{equation*}

\begin{Definition}
If the ideal J is closed with respect to the Poisson bracket
\begin{equation}\label{eq: coisotropic ideal}
\{J,J\}\subset {J} ,
\end{equation}
so that $\Gamma$ is a coisotropic submanifold of $R^{2n}$ ,
the functions $C_{jk}^{l}$ of the deformation parameters $x_{j}$
are said to define a coisotropic deformation of the associative
algebra.
\end{Definition}

Owing to the identity

\begin{eqnarray} \label{Schouten identity}
\{ f_{jk},f_{lr} \} &=&\sum_{m=1}^{n}\left[ C,C \right] _{jklr}^{m}p_{m}  \nonumber\\
&&+\sum_{m=1}^{n} \frac{\partial C_{lr}^{m}}{\partial {x_{k}}}f_{jm}+
\sum_{m=1}^{n} \frac{\partial C_{lr}^{m}}{\partial {x_{j}}}f_{km} \nonumber \\
&&-\sum_{m=1}^{n} \frac{\partial C_{jk}^{m}}{\partial {x_{r}}}f_{lm}
-\sum_{m=1}^{n} \frac{\partial C_{jk}^{m}}{\partial {x_{l}}}f_{rm}  ,
\end{eqnarray}
defining the Schouten-type bracket
\begin{eqnarray} \label{Schouten}
\left[ C,C \right] _{jklr}^{m}&=&\sum_{s=1}^{n} \left(
C_{sj}^{m}\frac{\partial C_{lr}^{s}}{\partial {x}_{k}}+
C_{sk}^{m}\frac{\partial C_{lr}^{s}}{\partial {x}_{j}}-
C_{sr}^{m}\frac{\partial C_{jk}^{s}}{\partial {x}_{l}} \right. \nonumber \\
&& \left. -C_{sl}^{m}\frac{\partial C_{jk}^{s}}{\partial {x}_{r}}+
\frac{\partial C_{jk}^{m}}{\partial {x}_{s}} C_{lr}^{s}-
\frac{\partial C_{lr}^{m}}{\partial {x}_{s}} C_{jk}^{s}  \right),
\end{eqnarray}
one readily sees that the structure constants define a coisotropic
deformation if and only if they satisfy the system of partial
differential equations
\begin{equation} \label{eq: coisotropy equations}
\left[ C,C \right] _{jklr}^{m}=0
\end{equation}
for any value of the indices $(j,k,l,m,r)$ ranging from zero to
the dimension of the algebra. This system and the associativity
conditions~(\ref{associativity equations}) are central in our
approach. For this reason they will be afterwards referred to as
the \textit{central system} of the theory of coisotropic
deformations.

The thesis of the present paper is that the central system
conceals a lot of interesting examples of dispersionless
integrable hierarchies of soliton theory, and that it
provides a new interesting route to understand their
integrability. This thesis is argued in six sections according
to the following plan:
\begin{itemize}
\item Sec.2 explains the origin of the new viepoint from the
theory of generalized dispersionless KP equations. In particular
it explains how to write these equations in the form of central
system. \item Sec.3 is a short study of the structure constants
associated with the dispersionless KP hierarchy, in search of
property of these constants which make the dKP equations
integrable. \item Sec.4 presents the first noticeable outcome of
the new approach. It shows that, for certain classes of algebras,
the associative conditions~(\ref{associativity equations}) and the
coisotropy conditions~(\ref{eq: coisotropy equations}) nicely
interact to produce the existence of a tau function seen as a
potential for the structure constants $C_{jk}^{l}$. It also shows
that the well-known dispersionless Hirota's bilinear equations are
nothing else than the associativity conditions~(\ref{associativity
equations}) written in terms of this potential. This result should
make easier the comprehension of the link between the Hirota's
equations and the associativity equations of Witten, Dijkgraaf,
Verlinde, Verlinde. \item Sec.5 is a brief study of the symmetry
properties of the central system, which has an infinite-
dimensional Abelian symmetry group. The study of the orbits of
this group in the space of structure constants allows to enlarge
the arsenal of interesting systems of structure constants at our
disposal, and to prove the invariance of the tau function under
the action of the symmetry group. \item Sec.6 presents the second
noticeable outcome of the new approach. The technique of quotient
algebras is used to glue together several copies of the algebras
associated with the dKP and dmKP hierarchies in such a way to
obtain new interesting examples of central systems. The class of
integrable equations covered by the new central systems is
sufficiently large to encompass the universal Whitham's hierarchy
of genus zero studied by Krichever. \item Sec.7 is a terse study
of coisotropic deformations from the viewpoint of the geometry of
the submanifolds $\Gamma$ previously introduced.
\end{itemize}
The paper ends with the indication of a few further possible
developments, and with two Appendices containing the details of
the computations presented along the paper.

\section{Introduction to the idea of coisotropic deformations}

The examples of dispersionless KP and mKP
equations are well suited to illustrate
the connection between coisotropic deformations
and integrable hierarchies. In this section we
recast these equations as a central system for
the structure constants of a simple commutative
associative algebra with unity.

In the early eighties, when the interest for the dispersionless
limit of soliton equations increased (see
e.g.~\cite{Za1}-\cite{Kri4}), it was well-known that the standard
dKP and dmKP equations arise as compatibility conditions of the
linear problem (see e.g.~\cite{ZMNP,AC})
\begin{eqnarray*}
\frac{\partial {\psi}}{\partial {x_{2}}} &=&
\frac{{\partial}^{2}{\psi}}{\partial {x_{1}^{2}}} +u_{1}(x_{1},x_{2},x_{3})\frac{\partial {\psi}}{\partial {x_{1}}}+u_{0}(x_{1},x_{2},x_{3})\psi\\
\frac{\partial {\psi}}{\partial {x_{3}}} &=& \frac{{\partial}^{3}{\psi}}{\partial {x_{1}^{3}}}
+v_{2}(x_{1},x_{2},x_{3})\frac{{\partial}^{2}{\psi}}{\partial {x_{1}^{2}}}+
v_{1}(x_{1},x_{2},x_{3})\frac{\partial {\psi}}{\partial {x_{1}}}+v_{0}(x_{1},x_{2},x_{3})\psi . \\
\end{eqnarray*}
It was therefore natural to assume that their dispersionless
limits are the compatibility conditions of the pair of time
dependent Hamilton-Jacobi equations
\begin{gather*}
\label{HJ}\begin{aligned} \frac{\partial S}{\partial x_{2}} &=
\left(\frac{\partial S}{\partial x_{1}} \right)^{2} + u_{1}
\left(x_{1},x_{2},x_{3} \right)
\frac{\partial S}{\partial x_{1}} + u_{0} \left(x_{1},x_{2},x_{3} \right), \\
\frac{\partial S}{\partial x_{3}} &= \left(\frac{\partial
S}{\partial x_{1}} \right)^{3} + v_{2} \left(x_{1},x_{2},x_{3}
\right) \left(\frac{\partial S}{\partial x_{1}} \right)^{2} +
v_{1} \left(x_{1},x_{2},x_{3} \right) \frac{\partial S}{\partial
x_{1}} + v_{0} \left(x_{1},x_{2},x_{3} \right)
\end{aligned}
\end{gather*}
The compatibility conditions were obtained, as in the dispersive
case, by imposing the equality of the second-order mixed
derivatives of the function $S(x_{1}$, $x_{2}$, $x_{3})$, and the
result was the following system of four partial differential
equations
\begin{gather}
\label{compatibility2}
\begin{aligned}
&\frac{\partial }{\partial {x_{1}}}(2v_{2}-3u_{1})=0 ,  \\
&\frac{\partial}{\partial {x_{1}}}(2v_{1}-3u_{0}) -
\frac{\partial {v_{2}}}{\partial {x_{2}}}+u_{1}\frac{\partial {v_{2}}}{\partial {x_{1}}}-2v_{2}\frac{\partial {u_{1}}}{\partial {x_{1}}} =0 ,\\
&2\frac{\partial {v_{0}}}{\partial {x_{1}}}- \frac{\partial
{v_{1}}}{\partial {x_{2}}}-2v_{2}\frac{\partial {u_{0}}}{\partial
{x_{1}}}-v_{1}\frac{\partial {u_{1}}}{\partial {x_{1}}}+
\frac{\partial {u_{1}}}{\partial {x_{3}}}=0  , \\
&-\frac{\partial {v_{0}}}{\partial {x_{2}}}+
u_{1}\frac{\partial {v_{1}}}{\partial {x_{1}}}-v_{1}\frac{\partial {u_{0}}}{\partial {x_{1}}}+\frac{\partial {u_{0}}}{\partial {x_{3}}} =0 . \\
\end{aligned}
\end{gather}
From it the dKP  and dmKP equations \cite{Kup1} are obtained
by setting
\begin{equation*}
u_{1}=0  \qquad  v_{2}=0  \qquad v_{1}=3/2u_{0} ,
\end{equation*}
and
\begin{displaymath}
u_{0}=0  \qquad  v_{0}=0  \qquad v_{2}=3/2u_{1}
\end{displaymath}
respectively. This selection of particular classes of
equations by means of suitable additional constraints
was called "gauge fixing".

Looking critically at this procedure, one may remark that the
method of compatibility conditions leaves a wide freedom in the
choice of the form of the auxiliary problem, allowing to
substitute the above pair of Hamilton-Jacobi equations by any
equivalent system of partial differential equations, without that
this change does modify the compatibility conditions and hence the
equations one is interested in. This inherent freedom is poorly
reflected by the above approach. For this reason we prefer to
follow a different route, by replacing the above pair of Hamilton-
Jacobi equations with the polynomial ideal
\begin{displaymath}
J=<h_{2},h_{3}>
\end{displaymath}
generated by the pair of Hamiltonian functions
\begin{eqnarray*}
h_{2} &=& -p_{2}+p_{1}^{2}+u_{1}(x_{1},x_{2},x_{3})p_{1}+u_{0}(x_{1},x_{2},x_{3}), \\
h_{3} &=&
-p_{3}+p_{1}^{3}+v_{2}(x_{1},x_{2},x_{3})p_{1}^{2}+v_{1}(x_{1},x_{2},x_{3})p_{1}+v_{0}(x_{1},x_{2},x_{3})    .\\
\end{eqnarray*}
The use of this ideal allows to account simultaneously for all the
possible forms of the auxiliary problems. The compatibility
conditions as easily recovered as the conditions saying that the
ideal $J$ is closed with respect to the classical Poisson bracket
in $R^{6}$. This classical viewpoint throws us immediately into
the theory of coisotropic deformations. To arrive to the central
system, it is sufficient to follow the construction of the dKP and
dmKP hierarchies.

The first higher equation in the dKP and dmKP hierarchies is
defined by the compatibility conditions expressing the closure
with respect to the classical Poisson bracket in $R^{8}$ of the
polynomial ideal generated by the  three Hamiltonians
\begin{eqnarray*}
h_{2} &=& -p_{2}+p_{1}^{2}+u_{1}(X)p_{1}+u_{0}(X) \\
h_{3} &=& -p_{3}+p_{1}^{3}+v_{2}(X)p_{1}^{2}+v_{1}(X)p_{1}+v_{0}(X)  \\
h_{4} &=& -p_{4}+p_{1}^{4}+w_{3}(X)p_{1}^{3}+w_{2}(X)p_{1}^{2}+w_{1}(X)p_{1}+w_{0}(X)  , \\
\end{eqnarray*}
where $X=(x_{1},x_{2},x_{3},x_{4})$ is the new set of coordinates.
Continuing the process, at each step
one introduces a new pair of coordinates $(x_{j},p_{j})$ and a new
Hamiltonian
\begin{equation*}
h_{j} = -p_{j} + P_{j}(p_{1}) \qquad for\quad  j\geq{2} ,
\end{equation*}
where $P_{j}$ is a monic polynomial of degree $j$ .
The sequence of polynomials $P_{j}$, defining the hierarchy,
does not contain the polynomials $P_{0}=1$ and $P_{1}=p_{1}$,
and therefore does not form a basis of the ring of polynomials.
To remedy this defect without changing the system of
compatibility conditions, it is expedient to introduce
four new coordinates $(x_{0},p_{0},x,p)$ and two new
Hamiltonians
\begin{equation*}
h_{0}=-p_{0}+1 \qquad  h_{1}=-p_{1}+p  ,
\end{equation*}
and to regard the polynomials $P_{j}$ as polynomials in
$p$ rather than in $p_{1}$.
The ideal $K$ generated by the
Hamiltonians
\begin{equation}\label{eq: hamiltonians}
h_{j} = -p_{j} + P_{j}(p)  \qquad for \quad  j\geq{0}
\end{equation}
has the same closure conditions of the
previous ideal $J$. Indeed the conditions
\begin{equation*}
\{h_{0},K\} \subset{K}  \qquad  \{h_{1},K\}\subset{K}
\end{equation*}
simply entail that $x_{0}$ is a cyclic coordinate,
and that $x$ appears always in the form $x+x_{1}$.
By this property the remaining closure conditions
\begin{equation*}
\{h_{j},K\}\subset{K}
\end{equation*}
give exactly the compatibility conditions of $J$.

The advantage of the completion of the basis is that
one may now consider the structure constants $C_{jk}^{l}(X)$
defined by
\begin{equation}\label{eq: table of multiplication}
P_{j}(p)P_{k}(p) = \sum_{l=0}^{l=j+k} C_{jk}^{l}(X)P_{l}(p)  .
\end{equation}
By construction they depend polynomially
on the coefficients of the Hamiltonians $h_{j}$, and
therefore they also satisfy a system of partial
differential equations. The main task of this section
is to identify these equations.

If one tries to attack the problem directly, by using
the explicit form of the dKP and dmKP equations, one may
get easily lost. The best strategy
is to use the properties of
the ideal $K$, and to notice that the double sequence of
polynomials
\begin{equation*}
f_{jk} = -p_{j}p_{k} +\sum_{l\geqq{0}}^{}C_{jk}^{l}(X)p_{l}
\end{equation*}
belong to $K$ since
\begin{equation}
f_{jk} = h_{j}h_{k} -\sum_{l\geqq{0}}^{}C_{jk}^{l}(X)h_{l}+p_{j}h_{k}+p_{k}h_{j} ,
\end{equation}
as one can easily check by using the definition
of the Hamiltonians $h_{j}$ .
For the condition $\{K,K\}\subset{K}$, each Poisson
bracket $\{ f_{lm},f_{pq} \}$ belongs to the ideal, and
therefore the structure constants $ C_{jk}^{l}$ satisfy
the equations $[C,C]_{jklr}^{m}=0$ as explained in the
Introduction. Since they obviously satisfy also the
associativity conditions, it is proved that the
structure constants of the dKP and dmKP equations satisfy
the central system.

To understand how the dKP and dmKP equations are sitting
inside this system, it is worth at this point to give a
closer look at the huge set of partial differential
equations $[C,C]_{jklr}^{m}=0$. Not to be lost, it is
convenient to fix at first the values of the indices
$(j,k,l,r)$. In this way one may
exploit the condition
\begin{equation}\label{triangularity}
C_{jk}^{l}=0 \qquad \textup{for} \qquad l>j+k
\end{equation}
to reduce drastically the number of equations.
For instance, for $j=k=l=1$ and $r=2$,
the infinite sequence of equations
\begin{gather*}
\begin{aligned}
\sum_{s\geq{0}} \left( C_{12}^{s}\frac{\partial {}}{\partial
{x_{s}}}C_{11}^{m}- C_{11}^{s}\frac{\partial {}}{\partial
{x_{s}}}C_{12}^{m}- C_{s2}^{m}\frac{\partial {}}{\partial
{x_{1}}}C_{11}^{s}- C_{s1}^{m}\frac{\partial {}}{\partial
{x_{2}}}C_{11}^{s} \right.\\
\left.+ C_{1s}^{m}\frac{\partial {}}{\partial {x_{1}}}C_{12}^{s}+
C_{1s}^{m}\frac{\partial {}}{\partial {x_{1}}}C_{12}^{s} \right)=
0 ,
\end{aligned}
\end{gather*}
contracts to four equations
\begin{gather}
\label{contraction}
\begin{aligned}
\sum_{s=0}^{s=3} \left(C_{12}^{s}\frac{\partial {}}{\partial
{x_{s}}}C_{11}^{m}- C_{11}^{s}\frac{\partial {}}{\partial
{x_{s}}}C_{12}^{m}- C_{s2}^{m}\frac{\partial {}}{\partial
{x_{1}}}C_{11}^{s}- C_{s1}^{m}\frac{\partial {}}{\partial
{x_{2}}}C_{11}^{s} \right. \\
\left.+ C_{1s}^{m}\frac{\partial {}}{\partial {x_{1}}}C_{12}^{s}+
C_{1s}^{m}\frac{\partial {}}{\partial {x_{1}}}C_{12}^{s} \right) =
0 \quad m=0,1,2,3
\end{aligned}
\end{gather}
owing to the triangularity condition~(\ref{triangularity}). In the
same way the choice $j=k=l=1$ and $r=3$ gives rise to five
equations, and so on.

The next question is to give these equations a sense. The
way is to have recourse to the table of multiplication
\begin{gather*}
\begin{aligned}
P_{1}P_{1} &= P_{2}-u_{1}P_{1}-u_{0}P_{0},  \\
P_{1}P_{2} &= P_{3}+(u_{1}-v_{2})P_{2}+(u_{0}-u_{1}^{2}+u_{1}v_{2}-v_{1})P_{1}+(v_{2}u_{0}-v_{0}-u_{1}v_{0})P_{0}, \\
P_{1}P_{3} &=
P_{4}+(v_{2}-w_{3})P_{3}+(v_{1}-w_{2}-v_{2}^{2}+w_{3}v_{2})P_{2}\\
&+(v_{2}^{2}u_{1}-u_{1}v_{2}w_{3}-v_{1}v_{2}+v_{1}w_{3}-u_{1}v_{1}+u_{1}w_{2}+v_{0}-w_{1})P_{1}\\
&-(w_{0}+v_{0}v_{2}+v_{0}w_{3}+u_{0}v_{1}-u_{0}w_{2})P_{0},  \\
P_{2}P_{2} &=
P_{4}+(2u_{1}-w_{3})P_{3}+(2u_{0}-w_{2}+u_{1}^{2}-2u_{1}v_{2}
+w_{3}v_{2})P_{2}\\
&+(2u_{1}^{2}v_{2}-2u_{1}v_{1}-u_{1}v_{2}w_{3}+v_{1}w_{3}-u_{1}^{3}+u_{1}w_{2}-w_{1})P_{1}\\
&+
(w_{0}+u_{0}w_{2}-u_{0}^{2}-u_{0}u_{1}^{2}+v_{0}w_{3}-2u_{1}v_{0}-u_{0}v_{2}w_{3}2u_{0}u_{1}v_{2})P_{0}  \\
\end{aligned}
\end{gather*}
of the polynomials $P_{j}$. From it one may read the
structure constants $ C_{jk}^{l}$ as functions of the
coefficients of the Hamiltonians $h_{j}$. For instance
\begin{eqnarray*}
C_{12}^{2} &=& u_{1}-v_{2}, \\
C_{12}^{1} &=& u_{0}-v_{1}-u_{1}^{2}+u_{1}v_{2},  \\
C_{12}^{0} &=& u_{0}v_{2}-v_{0}-u_{1}v_{0}  .\\
\end{eqnarray*}

The insertion of these expressions into the first
fragment~(\ref{contraction}) of the central system allows to
realize that these equations coincide with the set of four
compatibility conditions~(\ref{compatibility2}) used to defined
the dKP and dmKP equations. Similarly, one may check that the
fragment corresponding to $j=k=l=1$ and $r=3\;$ coincides with the
compatibility conditions defining the next members of the dKP and
dmKP hierarchies. In general one may prove that for $j=k=l=1$ and
$r$ arbitrary one obtains the full dKP and dmKP hierarchies.
Equivalently the same equations can be obtained by imposing the
vanishing of the Poisson brackets $\{f_{11},f_{1r}\}$ on the
submanifold $f_{1k}=0$, for $k=1,\ldots,r$.

At the end of this introductory discussion, it is interesting to
emphasize again that the discovery of the second interpretation of
the dispersionless KP hierarchy, as coisotropic deformation of the
structure constants of a specific associative algebra, is the
outcome of the replacement of the scheme of zero-curvature
representations with the scheme of coisotropic ideals, imposed by
the desire of making the theory covariant with respect to all
allowed changes of the auxiliary problems. The latter scheme
encompasses the former one, and allows to understand that there is
a unique mechanism behind the different representations of the
dispersionless KP hierarchy. Each representation is the expression
of the coisotropy of the ideal $K$ in a different system of
generators. By choosing the system of generators $h_{j}$ one
obtains the standard zero-curvature representation. By choosing
instead the system of generators $f_{jk}$ one arrives to see them
as equations controlling the evolution in time of the structure
constants of a certain associative algebra. This freedom in the
choice of the representation is a powerful tool for the study of
the integrable hierarchies. Soon we shall see that it allows to
account very easily for the passage from the zero-curvature
representation to the Hirota's representation of the dKP
hierarchy.

\section{The structure constants of dKP theory}

The plan of this section is to study the form of
the structure constants associated with the dispersionless KP
hierarchy. According to the viewpoint of the previous section,
this hierarchy is defined as the system of coisotropy conditions of
the submanifold $\Gamma$ formed by the zeroes of the Hamiltonian
functions
\begin{equation}
h_{j} = -p_{j}+\sum_{l=0}^{l=j-1}u_{jl}p_{1}^{l} , \nonumber
\end{equation}
and its structure constants are defined by the table of
multiplication
\begin{equation}
(h_{j}+p_{j})(h_{k}+p_{k}) = \sum_{l=1}^{n}C_{jk}^{l}(h_{l}+p_{l}) . \nonumber
\end{equation}
The point to be noticed is that the coisotropy conditions split in
two classes. Certain among them have the form of algebraic
constraints on the coefficients of the Hamiltonian functions, and
the question is to see the effect of these constraints on the form
of the structure constants $C_{jk}^{l}$. To have a reasonable
control of the question it is sufficient to consider the first few
Hamiltonians
\begin{eqnarray*}
h_{2} &=& -p_{2}+p_{1}^{2}+u_{0}(X), \\
h_{3} &=& -p_{3}+p_{1}^{3}+v_{1}(X)p_{1}+v_{0}(X),  \\
h_{4} &=& -p_{4}+p_{1}^{4}+w_{2}(X)p_{1}^{2}+w_{1}(X)p_{1}+w_{0}(X),  \\
h_{5} &=& -p_{5}+p_{1}^{5}+z_{3}(X)p_{1}^{3}+z_{2}(X)p_{1}^{2}+z_{1}(X)p_{1}+z_{0}(X) ,\\ \nonumber
\end{eqnarray*}
and the algebraic constraints
\begin{gather*}
\begin{aligned}
\label{compatibility conditions}
v_{1}&=3/2u_{0}, \\
w_{2}&=2u_{0},\\
w_{1}&=4/3v_{0}, \\
z_{3}&=5/2u_{0}, \\
z_{2}&=5/3v_{0}, \\
z_{1}&=5/4 w_{0}+ 5/8 u_{0}^{2},
\end{aligned}
\end{gather*}
 originating from the coisotropy conditions of the ideal $J$
generated by them. These constraints may be encoded into the
definition of a special class of polynomials, henceforth called
\textit{Faa' di Bruno polynomials}. The first six polynomials are
\begin{eqnarray*}
P_{0}(p)  &=& 1,  \\
P_{1}(p)  &=& p,  \\
P_{2}(p)  &=& p^{2}+u_{0},  \\
P_{3}(p)  &=& p^{3}+3/2u_{0}p+v_{0},  \\
P_{4}(p)  &=& p^{4}+2u_{0}p^{2}+4/3v_{0}p+w_{0}, \\
P_{5}(p)  &=& p^{5}+5/2u_{0}p^{3}+5/3v_{0}p^{2}+(5/4w_{0}+5/8u_{0}^{2})+z_{0}. \\
\end{eqnarray*}
Our interest is in their table of multiplication. The simplest
part is
\begin{eqnarray*}
P_{1}P_{1} &=& P_{2}-[u_{0}]P_{0},  \\ \nonumber P_{1}P_{2} &=&
P_{3}-[1/2u_{0}]P_{1}-[v_{0}]P_{0},  \\  \nonumber P_{1}P_{3} &=&
P_{4}-[1/2u_{0}]P_{2}-[1/3v_{0}]P_{1}-[w_{0}-1/2u_{0}^{2}]P_{0}, \\
\nonumber P_{1}P_{4} &=&
P_{5}-[1/2u_{0}]P_{3}-[1/3v_{0}]P_{2}-[1/4w_{0}-1/8u_{0}^{2}]P_{1}
-[z_{0}-5/6u_{0}v_{0}]P_{0} .\\  \nonumber
\end{eqnarray*}

The coefficients of these equations are the footprints
of the structure constants we are looking for. One
may notice that each coefficient becomes stationary exactly
one step after it appears for the first time. To account for
this phenomenon it is useful to rewrite the table in the form
\begin{gather*}
\begin{aligned}
P_{1}P_{1} = P_{2}&-[1/2u_{0}]P_{0}-[1/2u_{0}]P_{0},  \\ \nonumber
P_{1}P_{2} =
P_{3}&-[1/2u_{0}]P_{1}-[1/3v_{0}]P_{0}-[2/3v_{0}]P_{0},  \\
\nonumber P_{1}P_{3} =
P_{4}&-[1/2u_{0}]P_{2}-[1/3v_{0}]P_{1}-[1/4w_{0}-1/8u_{0}^{2}]P_{0}\\
&-[3/4w_{0}-3/8u_{0}^{2}]P_{0},
\\  \nonumber P_{1}P_{4} =
P_{5}&-[1/2u_{0}]P_{3}-[1/3v_{0}]P_{2}-[1/4w_{0}-1/8u_{0}^{2}]P_{1}\\
&-[1/5z_{0}-1/6u_{0}v_{0}]P_{0}-[4/5z_{0}-2/3u_{0}v_{0}]P_{0},
\end{aligned}
\end{gather*} showing that
\begin{equation}
P_{1}P_{j} = P_{j+1}+\sum_{l=1}^{j}H_{l}^{1}P_{j-l}+\sum_{l=1}^{1}H_{l}^{j}P_{1-l} \nonumber
\end{equation}
for two suitable sequences of constants $(H_{l}^{1},H_{1}^{l})$.
Things go similarly for the second table of multiplication
\begin{eqnarray*}
P_{2}P_{1} &=&
P_{3}-([2/3v_{0}]P_{0})-([1/2u_{0}]P_{1}+[1/3v_{0}]P_{0}),  \\
\nonumber P_{2}P_{2} &=&
P_{4}-([2/3v_{0}]P_{1}+[1/2w_{0}-1/2u_{0}^{2}]P_{0})-([2/3v_{0}]P_{1}
+[1/2w_{0}-1/2u_{0}^{2}]P_{0}),  \\  \nonumber P_{2}P_{3} &=&
P_{5} -([2/3v_{0}]P_{2}+[1/2w_{0}-1/2u_{0}^{2}]P_{1}+[****]P_{0})
-([***]P_{1}+[***]P_{0}),  \\  \nonumber
\end{eqnarray*}
showing that
\begin{equation}
P_{2}P_{j} = P_{j+2}+\sum_{l=1}^{j}H_{l}^{2}P_{j-l}+\sum_{l=1}^{2}H_{l}^{j}P_{2-l}  \nonumber
\end{equation}
for two new sequences of constants $(H_{l}^{2},H_{2}^{l})$. In general
one may expect (and prove subsequently) that
\begin{equation}
P_{k}P_{j} = P_{k+j}+\sum_{l=1}^{j}H_{l}^{k}P_{j-l}+\sum_{l=1}^{k}H_{l}^{j}P_{k-l}  .
\end{equation}
By this formula we have reached our aim, discovering that
the structure constants of the Faa' di Bruno polynomials
have the form
\begin{equation}
\label{structure_constants} C_{kj}^{l} =
\delta_{k+j}^{l}+H_{j-l}^{k}+H_{k-l}^{j}  .
\end{equation}
where $\delta_{k}^{l}$ is the Kronecker symbol. This formula is
the starting point of the algebraic analysis of the Hirota's
bilinear formulation of the dispersionless KP hierarchy.

\section{Tau function and Hirota's equations.}

The important question which remains unanswered
at the bottom of the algebraic approach is to
understand the link of coisotropic deformations
to integrability. So far we have realized that
certain integrable partial differential equations
may be written in the form of central system,
but we do not yet know if all that
has a sense. In this section we point out a second
occurrence that partly answers the previous question.

We assume that the structure constants have the form suggested by
the examination of the dKP hierarchy, without specifying the
coefficients $H_{j}^{k}$, and we investigate the implications of
the process of coisotropic deformation for this class of algebras.
Our purpose is to show that the associativity conditions
(\ref{associativity equations}) and the coisotropy conditions
(\ref{eq: coisotropy equations}) cooperate to produce the
existence of a tau function, which is the first sign of the
integrability of the given set of partial differential equations.

The first step is to implement the associativity conditions on the
coefficients $H_{j}^{k}$.  By direct substitution of
expressions~(\ref{structure_constants}) into
equations~(\ref{associativity equations}), and by the use of the
identity
\begin{equation}\label{identity}
\sum_{l=n-1}^{k-1}H_{k-l}^{i}H_{l-n}^{m}=
\sum_{l=n-1}^{k-1}H_{k-l}^{m}H_{l-n}^{i}  ,
\end{equation}
one is led to the system of equations
\begin{gather}
\label{associativity0}
\begin{aligned}
&H_{i+k-n}^{m}+H_{m-n}^{i+k}-H_{m+k-n}^{i}-H_{m+k}^{i-n}\\
&+\theta(n-m)H_{i+m-n}^{k}+\theta(n-m)H_{k+m-n}^{i}\\
&-\theta(n-i)H_{i+m-n}^{k}-\theta(n-i)H_{i+k-n}^{i}\\
&+\sum_{l=n-1}^{i-1}H_{i-l}^{k}H_{l-n}^{m}+\sum_{l=1}^{i-1}H_{i-l}^{k}H_{m-n}^{l}\\
&+\sum_{l=1}^{n-1}H_{k-l}^{i}H_{m-n}^{l}-\sum_{l=n-1}^{m-1}H_{m-l}^{k}H_{l-n}^{i}\\
&-\sum_{l=1}^{m-1}H_{m-l}^{k}H_{i-n}^{l}
-\sum_{l=1}^{k-1}H_{k-l}^{m}H_{i-n}^{l} = 0 ,
\end{aligned}
\end{gather}
 where $\theta(n)=1$ or $\theta(n)=0$ according if $n\geqslant{0}$
or $n<0$. However these equations are not all independent. The
analysis of Appendix A allows to reduce the number of
equations, and shows that the structure constants $C_{jk}^{l}$
obey the associative conditions if and only if the bracket
\begin{equation*}
\left [ H,H \right ]_{ikm}:= H_{m}^{i+k}-H_{m+k}^{i}-H_{i+m}^{k}
+\sum_{l=1}^{i-1}H_{i-l}^{k}H_{m}^{l}+\sum_{l=1}^{k-1}H_{k-l}^{i}H_{m}^{l}
-\sum_{l=1}^{m-1}H_{m-l}^{k}H_{l}^{i}
\end{equation*}
vanishes identically for any choice of the indices
$(i,k,m)\in{N}$. An interesting consequence of this result can be
drawn by contraction. Indeed one may check that the equations
\begin{equation}\label{quadrics}
\sum_{m+k=p} [ H,H ]_{ikm} = 0
\end{equation}
and the use of the identities~(\ref{identity}) lead to the
contracted identities
\begin{equation*}
pH_{p}^{i} = H_{p}^{i}+\sum_{k=1}^{p-1}(H_{k}^{i+p-k}-H_{i+p-k}^{k})
-\sum_{l=1}^{i-1}\sum_{k=1}^{p-1}H_{l}^{p-k}H_{k}^{i-l}  ,
\end{equation*}
entailing the useful symmetry relations
\begin{equation}
\label{symmetry relations_bis} pH_{p}^{i} = iH_{i}^{p}  .
\end{equation}
As we shall see in a moment, this remarkable outcome
of the associativity conditions is at the basis of
Hirota's formulation.

Next one has to implement the coisotropy conditions
\label{coisotropy conditions}. In terms of the coefficients
$H_{k}^{i}$ the ensuing equations are, at first sight, rather
complicated, and they are presented in Appendix A (as a particular
case of equation (\ref{quarta}) ). However a closer scrutiny shows
that they are simplified drastically on account of the
associativity conditions just obtained. Indeed in Appendix A it is
proved that the coisotropy conditions may be reduced to the
equations
\begin{equation}
\frac{\partial [ H,H ]_{l,n,i-j}}{\partial x_{k}} + \frac{\partial
[ H,H ]_{l,n,k-j}}{\partial x_{i}} - \frac{\partial [ H,H
]_{i,k,l-j}}{\partial x_{n}} - \frac{\partial [ H,H
]_{i,k,n-j}}{\partial x_{l}}  = 0 , \nonumber
\end{equation}
which are automatically fulfilled owing to the
associativity conditions, and to the \textit{linear}
equations
\begin{equation}
\frac{\partial H_{p}^{i}}{\partial x_{l}} =
\frac{\partial H_{p}^{l}}{\partial x_{i}}  . \nonumber
\end{equation}

So, to summarize, the associativity and coisotropy conditions in
the case of the structure constants of the
form~(\ref{structure_constants}) are together equivalent to the
set of quadratic algebraic equations
\begin{equation}\label{associativity2}
[ H,H ]_{ikl}=0 ,
\end{equation}
entailing the symmetry conditions
\begin{equation}\label{symmetry relations}
pH_{p}^{i} = i H_{i}^{p}  ,
\end{equation}
and to the set of linear differential equations
\begin{equation}\label{exactness}
\frac{\partial H_{p}^{i}}{\partial x_{l}} =
\frac{\partial H_{p}^{l}}{\partial x_{i}}
\end{equation}
having the form of a system of conservation laws. The
equations~(\ref{associativity2}) and~(\ref{exactness}) give the
specific form of the \textit{central system} of the dispersionless
KP hierarchy. It encodes all the informations about the hierarchy.
In particular it entails that for any solution of the central
system one has
\begin{equation}\label{brackets}
\{ f_{ik},f_{ln} \}=\sum_{s,t \geq 1  } K_{ikln}^{st} f_{st}
\end{equation}
where
\begin{equation*}
K_{ikln}^{st}= \left (\delta_{it}\frac {\partial }{\partial
x_{k}}+\delta_{kt}\frac{\partial }{\partial x_{i}} \right)
(H_{l-s}^{n}+H_{n-s}^{l})- (\delta_{nt}\frac{\partial}{\partial
x_{l} }+\delta_{lt}\frac{\partial}{\partial x_{n}
})(H_{i-s}^{k}+H_{k-s}^{i})  .
\end{equation*}
From this formula one sees that the Hamiltonians $f_{jk}$ of the
dispersionless KP hierarchy form a Poisson algebra. The above
central system can be seen also as the dispersionless limit of the
central system of the full dispersive KP hierarchy~\cite{Ma1}.

There are presently two strategies to decode the informations
contained into the central system. According to the first
strategy, one first tackles the associativity
conditions~(\ref{associativity2}), noticing that they allow to
compute the coefficients $(H_{k}^{2},H_{k}^{3},\ldots)$ as
polynomial functions of $H_{k}^{1}$. For instance, the symmetry
conditions
\begin{equation*}
H_{1}^{2}=2H_{2}^{1}  \qquad   H_{1}^{3} = 3H_{3}^{1}
\end{equation*}
give $(H_{1}^{2},H_{1}^{3})$, and then the
condition~(\ref{associativity2}) with $i=k=1$, $l=2$, i.e.
\begin{equation*}
H_{1}^{3}-H_{3}^{1}-H_{2}^{2}+H_{1}^{1}H_{1}^{1} = 0
\end{equation*}
gives $H_{2}^{2}$, and so forth. Renaming the free coefficients as
suggested by the table of multiplication of the previous section,
that is by setting
\begin{equation*} H_{1}^{1}=-1/2u_{0},  \qquad
H_{2}^{1}=-1/3v_{0},  \qquad H_{3}^{1}=-1/4w_{0}+1/8u_{0}^{2} ,
\end{equation*}
one gets
\begin{equation*}
H_{1}^{2}=-2/3v_{0},  \qquad  H_{2}^{2}=-1/2w_{0}+1/2u_{0}^{2},
\qquad H_{1}^{3}=-3/4w_{0}+3/8u_{0}^2.
\end{equation*}
At this point one may plug these expressions into the simplest
linear coisotropy conditions
\begin{equation*}
\frac{\partial H_{1}^{1}}{\partial x_{2} }-\frac{\partial
H_{1}^{2}}{\partial x_{1} }=0, \qquad \frac{\partial
H_{2}^{1}}{\partial x_{2} }-\frac{\partial H_{2}^{2}}{\partial
x_{1} }=0, \qquad
\frac{\partial H_{1}^{1}}{\partial x_{3} }-\frac{\partial H_{1}^{3}}{\partial x_{1} }=0 ,\\
\end{equation*}
arriving to the equations
\begin{align*}
\frac{\partial v_{0} }{\partial x_{1} } &= 3/4\frac{\partial u_{0} }{\partial x_{2} }, \\
\frac{\partial v_{0} }{\partial x_{2} } &= 3/2\frac{\partial w_{0}
}{\partial x_{1} }-3u_{0}\frac{\partial u_{0} }{\partial x_{1}},
\\
\frac{\partial u_{0} }{\partial x_{3} } &=
3/2\frac{\partial w_{0} }{\partial x_{1} }-3/2u_{0}\frac{\partial u_{0} }{\partial x_{1} }  .\\
\end{align*}
The elimination of $\frac{\partial w_{0} }{\partial x_{1} }$
leads finally to the dispersionless KP equation and to the
higher equations, if one insists enough in the computations.
By this strategy one come back to the hierarchy in its
standard formulation.

A simple inversion in the order in which the equations are
considered leads instead to the Hirota's formulation. It is enough
to remark that equations (\ref{exactness}) entail the existence of
a sequence of potentials $S_{m}$ such that
\begin{equation*}
H_{m}^{i} = \frac{\partial S_{m} }{\partial x_{i} }  .
\end{equation*}
Then the symmetry conditions (\ref{symmetry relations}), oblige
the potentials $S_{m}$ to obey the constraints
\begin{equation}
i\frac{\partial {S_{i}}}{\partial {x_{l}}}=l\frac{\partial {S_{l}}}{\partial {x_{i}}}  , \nonumber
\end{equation}
which in turn entail the existence of a superpotential
$F(x_{1},x_{2},\ldots)$ such that
\begin{displaymath}
S_{i}=-1/i\frac{\partial {F}}{\partial {x_{i}}},  \qquad
H_{m}^{i}=-1/m\frac{\partial^{2}{F}}{\partial {x_{i}}\partial
{x_{m}}}.
\end{displaymath}
This result provides a second parametrization of the structure
constants, after that described before. The insertion of the new
parametrization into the full set of associativity conditions
\label{associativity} finally leads to the system of equations
\begin{gather}
\begin{aligned}\label{Hirota}
&-\frac{1}{m}F_{i+k,m}+\frac{1}{m+k}F_{i,k+m}+\frac{1}{i+m}F_{k,i+m}\\
&+\sum_{l=1}^{i-1}\frac{1}{m(i-l)}F_{k,i-l}F_{l,m}+\sum_{l=1}^{k-1}\frac{1}{m(k-l)}F_{i,k-l}F_{l,m}
-\sum_{l=1}^{m-1}\frac{1}{i(m-l)}F_{k,m-l}F_{i,l}=0  ,
\end{aligned}
\end{gather}
where $F_{i,k}$ stands for the second-order derivative of $F$ with
respect to $x_{i}$ and $x_{k}$. They are equivalent to the
celebrated Hirota's bilinear equations for the tau function of the
dispersionless KP hierarchy (see e.g. \cite{TT1,TT2,Car1,Mar1}).
For instance, for $(i=k=1,m=2)$ or
 $(i=m=1,k=2)$ or  $(m=k=1,i=2)$ one obtains directly the first Hirota's
equation
\begin{equation}
-1/2F_{2,2}+2/3F_{1,3}-(F_{1,1})^{2}=0  . \nonumber
\end{equation}
For $(i=1,k=2,m=2)$ or $(i=2,k=1,m=2)$ or $(i=1,k=1,m=3)$ it
gives instead the second Hirota's equation
\begin{equation}
1/2F_{1,4}-1/3F_{2,3}-F_{1,1}F_{1,2}=0  . \nonumber
\end{equation}
The choices $(i=1,k=1,m=4)$ and $(i=1,k=2,m=3)$ lead to
the equations
\begin{equation}
-1/4F_{2,4}+2/5F_{1,5}-2/3F_{1,1}F_{1,3}-1/4(F_{1,2})^{2}=0  \nonumber
\end{equation}
and
\begin{equation}
-1/3F_{3,3}+1/5F_{1,5}+1/4F_{2,4}+1/3F_{1,1}F_{1,3}-1/2F_{1,1}F_{2,2}-1/2(F_{1,2})^{2}=0  \nonumber
\end{equation}
respectively, which do not separately coincide with
Hirota's bilinear equations, but which are together
equivalent to a pair of standard Hirota's equations of
the same weight. The process of identification
may be continued indefinitely.

The above result lends itself to several comments. The discovery
that the Hirota's equations of the dispersionless KP hierarchy are
the \textit{associativity equations} for the structure constants
(\ref{structure_constants}) immediately reminds the associativity
equations of Witten, Dijkgraaf, Verlinde, Verlinde~\cite{W,DVV},
first derived in the frame of two-dimensional topological field
theories and subsequently interpreted by Dubrovin \cite{Dub4} in
the frame of his theory of Frobenius's manifolds. In spite of the
ideological resemblance there are, however, essential differences
between the present approach and that of Frobenius manifolds
developped by Dubrovin. In the latter approach the structure
constants are derived from solutions of WDVV equations and verify
the associativity conditions~(\ref{associativity equations}) but
not the full central system. They define deformations of Frobenius
algebras, which are associative algebras endowed with a
nondegenerate symmetric form, and are attached to the tangent
spaces of a Frobenius manifold whose points are parametrized by
the deformation coordinates $x_{j}$. In the present approach the
structure constants are derived directly from the hierarchy and do
not presuppose the preliminary knowledge of the tau function.
Instead they serve to introduce the tau function, and not
viceversa as happens in the theory of Frobenius manifolds.
Furthermore they are used to define a coisotropic submanifold in
the cotangent  bundle. So the two approaches live in different
spaces. Furthermore, the Riemannian structure of the base space,
which is fundamental for Dubrovin, seems untied with the present
approach. Viceversa the symplectic structure of the cotangent
bundle, which is fundamental for us, seems do not play any role in
the Dubrovin's approach. The exact relation between these
different vision is therefore a complicated problem. We note,
however, that the functions $K_{ikln}^{st}$ in
formula~(\ref{brackets}) are linear combinations of third- order
derivatives ot the tau function $F$. For example,
\begin{equation}
\{ f_{11},f_{1n} \} = -2\sum_{s=1}^{s=n-1}( \frac{1}{n-s}
\frac{\partial^3 F}{{\partial x_{1} }^{2} \partial x_{n-s} })f_{1s} \quad n=2,3,\ldots . \nonumber
\end{equation}
It is, probably, not just a coincidence that third order
derivative appear  in both WDVV equations and as structure
constants in equations~(\ref{brackets}).

From another point of view, one may also consider the central
system as an interesting linearization of the dKP hierarchy. In
terms of the coefficients of the Hamiltonians $h_{j}$ this system
is just the dKP hierarchy of nonlinear PDE's. In terms of
variables $H_{k}^{i}$ the dKP hierarchy is represented by linear
exactness conditions on the family of quadrics~(\ref{quadrics}). A
linearization of nonlinear PDE's integrable by the inverse
scattering transform method in terms of appropriate variables (
inverse scattering data) is a common property of such equations
(see. e.g. \cite{ZMNP,AC}). In particular, in the papers
\cite{MT,Dei} it was shown that the Korteweg-de-Vries ,nonlinear
Schrodinger, sine-Gordon, and Toda lattice equations are
linearizable (become systems of free oscillators) on the
intersection of quadrics in phase space. The peculiarity of the
dispersionless hierarchies is that the family of quadrics is
defined by the associativity conditions for certain algebras. In
this vein, the associativity conditions also remind the relations
for Plucker's coordinates within the Sato's universal Grassmannian
approach to the full dispersive KP hierarchy~\cite{Sa}.

\section{The symmetry action}

So far we have considered a relatively narrow class
of algebras and of dispersionless integrable
equations. To enlarge this class we shall follow
two different routes, in this and the next
section, having recourse to the action of
a symmetry group and to a process of gluing.

The transformation of the structure constants
\begin{equation}
\widetilde{C}_{ab}^{d}(X)=
\sum_{a,b,l>0}C_{jk}^{l}(X)A_{a}^{j}(X)A_{b}^{k}(X)A_{l}^{d}(X) ,
\end{equation}
induced by a change of basis
\begin{equation}\label{change of basis}
\tilde{p}_a= \sum_{l>0} A_{a}^{l} (X) p_{l}
\end{equation}
depending on deformation parameters $X=(x_{1}, x_{2},\ldots)$,
obviously preserves the symmetry conditions~(\ref{commutativity
equations}) and the associativity conditions~(\ref{associativity
equations}), but violates in general the coisotropy
conditions~(\ref{Schouten}).

The class of transformations preserving the last
conditions can be worked out by turning to
the correspondence between deformation parameters
and generators, on one side, and coordinates and
momenta, on the other side, which has been used to
establish the Hamiltonian interpretation of the
process of coisotropic deformation. Shifting from
commutative algebra to Hamiltonian mechanics, one
then substitutes the above change of basis
in the associative algebra with the
transformation of coordinates
\begin{gather}
\label{transformation}
\begin{aligned}
\tilde{x}_{a} &= \sum_{l>0} \delta_{a}^{l} x_{l} \\
\tilde{p}_{a} &= \sum_{l>0} A_{a}^{l} (X) p_{l}
\end{aligned}
\end{gather}
on the symplectic manifold related to the algebra.
This change of coordinates acts on the Hamiltonians
$f_{jk}$ according to the transformation law
\begin{equation}
\tilde{f}_{ab}(X) = \sum_{j,k>0}A_{a}^{j}(X)A_{b}^{k}(X)f_{jk}(X),
\end{equation}
which shows that the functions $\tilde{f}_{ab}$ still belong to
the ideal $J$ generated by the functions $f_{jk}$. This ideal is
thus left invariant by the transformation~(\ref{transformation}),
but it looses in general the property of being closed with respect
to the classical Poisson bracket since the bracket is modified by
the coordinate transformation. So to preserve the coisotropy
conditions we need to restrict the change of basis in such a way
that the transformation~(\ref{transformation}) be canonical. This
is a severe restriction and the unique solution is
\begin{equation}
\tilde{p}_{a}=p_{a}+\frac{\partial {\phi}}{\partial {x_{l}}}p_{0}
\end{equation}
where $\phi$ is an arbitrary function of the deformation parameters.
The conclusion is that there exists an infinite-dimensional
Abelian symmetry group of the equations defining the coisotropic
deformations of associative algebras depending on
an arbitrary function $\phi$. This group is
the subject of this section.

The action of the group on the
space of structure constants is defined by

\begin{eqnarray}
\widetilde{C}_{jk}^{l} &=& C_{jk}^{l}+
\delta_{k}^{l}\frac{\partial \phi}{\partial x_{j}}+
\delta_{j}^{l}\frac{\partial \phi}{\partial x_{k}}, \\
\widetilde{C}_{jk}^{0} &=& C_{jk}^{0}+ \frac{ \partial
\phi}{\partial x_{j}}\frac{\partial \phi}{\partial x_{k}}-
\sum_{m>0} C_{jk}^{m} \frac{\partial {\phi}}{\partial x_{m}}
\end{eqnarray}
for $j,k,l>0$.
Therefore the equations of the orbit
passing through the point corresponding to the
structure constants of the dispersionless KP hierarchy are
\begin{equation}
\label{newstructureconstants} \widetilde{C}_{jk}^{l}=
\delta_{j+k}^{l}+\widetilde{H}_{k-l}^{j}+
\widetilde{H}_{j-l}^{k}+\delta_{0}^{l}\widetilde{D}_{jk} ,
\end{equation}
with the understanding that the coefficients are given by
\begin{gather*}
\begin{aligned}
\widetilde{H}_{k}^{i} &= H_{k}^{i},  \\
\widetilde{H}_{0}^{i} &= \frac{\partial {\phi}}{\partial {x_{i}}},   \\
\widetilde{H}_{k}^{0} &= \delta_{k}^{0}, \\
\widetilde{D}_{ik}    &= -\frac{\partial {\phi}}{\partial
{x_{i+k}}}- \frac{\partial {\phi}}{\partial {x_{i}}}\frac{\partial
{\phi}}{\partial {x_{k}}}-
\sum_{m=1}^{m=i-1}H_{i-m}^{k}\frac{\partial {\phi}}{\partial
{x_{m}}}-
\sum_{m=1}^{m=k-1}H_{k-m}^{i}\frac{\partial {\phi}}{\partial {x_{m}}},  \\
\widetilde{D}_{0k}    &= -\delta_{k}^{0} \\
\end{aligned}
\end{gather*}
for $i,k>0$. One may notice that the new structure constants
$\widetilde{C}_{jk}^{l}$ closely resemble the original ones, and
one may infer from this fact that they also derive from a tau
function. The first problem we are interested in is to inquiry how
the symmetry group acts at the level of tau functions.

The proof of the existence of a tau function $
\widetilde{F}(x_{1},x_{2},\ldots)$ for the new structure constants
$\widetilde{C}_{jk}^{l}$ strictly follows the pattern of the
previous section. The strategy is always to work out the
implications on the coeffifients $\widetilde{H}_{k}^{i}$ and
$\widetilde{D}_{ik}$ of the conditions of coisotropy and
associativity. In Appendix A it is shown that the conditions of
associativity give rise to two sets of constraints, affecting
separately the two families of coefficients. One set fixes the
form of the coefficients $ \widetilde{D}_{ik}$ according to
\begin{equation*}
\widetilde{D}_{ik} =
-\widetilde{H}_{0}^{i+k}+\widetilde{H}_{0}^{i}\widetilde{H}_{0}^{k}-
\sum_{l=1}^{l=k}\widetilde{H}_{k-l}^{i}\widetilde{H}_{0}^{l}-
\sum_{l=1}^{l=k}\widetilde{H}_{i-l}^{k}\widetilde{H}_{0}^{l}  ,
\end{equation*}
while the second set requires that the coefficients
$\widetilde{H}_{k}^{i}$ verify the already known conditions
\begin{equation*}
\widetilde{H}_{m}^{i+k}-\widetilde{H}_{m+k}^{i}-\widetilde{H}_{i+m}^{k}
+\sum_{l=1}^{i-1}\widetilde{H}_{i-l}^{k}\widetilde{H}_{m}^{l}+\sum_{l=1}^{k-1}\widetilde{H}_{k-l}^{i}\widetilde{H}_{m}^{l}
-\sum_{l=1}^{m-1}\widetilde{H}_{m-l}^{k}\widetilde{H}_{l}^{i} = 0
.
\end{equation*}
A closer scrutiny of these conditions shows that they do not
contain terms of the form $\widetilde{H}_{0}^{i}$, since these
terms cancel in pair, proving that the coefficients
$\widetilde{H}_{k}^{i}$, for $i,k>0$, satisfy again the symmetry
conditions
\begin{equation*}
p\widetilde{H}_{p}^{i} = i\widetilde{H}_{i}^{p}  .
\end{equation*}
Similarly the coisotropy conditions also split in two sets that,
owing to the associativity conditions, take the form
\begin{eqnarray*}
\frac{\partial {\widetilde{H}_{0}^{l}}}{\partial {x_{i}}} &=&
\frac{\partial {\widetilde{H}_{0}^{i}}}{\partial {x_{l}}},\\
\frac{\partial {\widetilde{H}_{p}^{l}}}{\partial {x_{i}}} &=&
\frac{\partial {\widetilde{H}_{p}^{i}}}{\partial {x_{l}}}
\end{eqnarray*}
for $i,l,p>0$. One infers from the last conditions the existence
of a sequence of potentials $\phi=\widetilde{S}_{0}$ and $
\widetilde{S}_{k}$, such that
\begin{equation*}
\widetilde{H}_{0}^{i}=\frac{\partial {\phi}}{\partial {x_{i}}},
\qquad \widetilde{H}_{k}^{i}=\frac{\partial
{\widetilde{S}_{k}}}{\partial {x_{i}}} .
\end{equation*}
By the same argument discussed in Sec.4, the potentials
$\widetilde{S}_{k}$ lead to the existence of a single function
$\widetilde{F}(x_{1},x_{2},\ldots)$  such that
\begin{equation*}
\widetilde{H}_{k}^{i}=-\frac{1}{k}\frac{\partial^{2}{\widetilde{F}}}
{\partial {x_{i}}\partial {x_{k}}}  .
\end{equation*}
The existence of the potential $\phi$  shows instead that the new
structure constants $\widetilde{C}_{jk}^{l}$ belong to the orbit
passing through the point representing the KP hierarchy. Due to
the invariance of the coefficients $\widetilde{H}_{k}^{i}$ along
the orbit ($i,h>0$), it turns out that the tau function is
invariant along the orbit. Thus this function is a property of the
orbit rather than of the points of the orbit.

The second problem of our concern is to investigate the properties
of a second remarkable point belonging to the orbit, whose definition
is suggested by the equations of the orbit. They entail, in particular,
that
\begin{equation*}
\widetilde{C}_{jk}^{0}=\widetilde{H}_{k}^{j}+
\widetilde{H}_{j}^{k}+\widetilde{D}_{jk} .
\end{equation*}
This formula shows that the structure constants
$\widetilde{C}_{jk}^{0}$ vary along the orbit, and therefore one
can ask if there exists a point on the orbit where these structure
constants vanish. The answer is affirmative owing to the first
half of the associativity conditions. Indeed to demand that
$\widetilde{C}_{jk}^{0}=0$ is equivalent to search a function
$\phi$ such that
\begin{equation*}
\frac{\partial {\phi}}{\partial {x_{i+k}}} +\frac{\partial
{\phi}}{\partial {x_{i}}}\frac{\partial {\phi}}{\partial {x_{k}}}
+\sum_{m=1}^{m=i-1}H_{i-m}^{k}\frac{\partial {\phi}}{\partial
{x_{m}}} +\sum_{m=1}^{m=k-1}H_{k-m}^{i}\frac{\partial
{\phi}}{\partial {x_{m}}} = H_{k}^{i}+H_{i}^{k}  ,
\end{equation*}
if one takes into account the invariance of the coefficients
$H_{k}^{i}$ along the orbit. The system just written is an
overdetermined system of nonlinear partial differential equations
on the single unknown function $\phi$ whose compatibility
conditions are exactly the first half of the associativity
conditions. Accordingly there exists a function $\phi$ which
allows to reach the desired point from the point corresponding to
dispersionless KP. It is readily seen that the new point
correspond to the structure constants of the dispersionless mKP
hierarchy. It is sufficient to interpret the last equations as the
hierarchy of dispersionless Miura transformations relating the
dispersionless KP and mKP hierarchies. This interpretation is
motivated by the remark that, at  the lowest level $i=k=1$, the
equation of the symmetry action has the form
\begin{equation*}
\frac{\partial {\phi}}{\partial {x_{2}}}+(\frac{\partial
{\phi}}{\partial {x_{1}}})^2= 2H_{1}^{1},
\end{equation*}
and therefore becomes
\begin{equation*}
u_{0}=1/2 \frac{\partial^{-1}}{\partial {x_{1}}}(\frac{\partial {u_{1}}}{\partial {x_{2}}})
-1/4u_{1}^{2}
\end{equation*}
upon the insertion of the standard parametrization
\begin{equation*}
u_{0}=-2H_{1}^{1},  \qquad u_{1}=-2H_{0}^{1}=-2\frac{\partial
{\phi}}{\partial {x_{1}}} .  \nonumber
\end{equation*}
The above equation is the dispersionless limit of the well-known
Miura transformation between the KP and mKP equations (see e.g.
\cite{Li,Ch}). The lesson is that the dispersionless Miura
transformation is just a particular instance of the symmetry
action generated by the special changes of basis~(\ref{change of
basis}) in the associative algebra. At the level of structure
constants the Miura transformations have therefore a very simple
meaning.

It remains to identify the meaning of the other points of the
orbits. A possible way is to carefully study the parametrization
of the structure constants $\widetilde{C}_{jk}^{l}$ obeying the
equations of the orbit. We omit this study and we limit ourselves
to give the final answer. As the reader may expect at this point,
one may prove that the structure constants of the orbit passing
through the dKP hierarchy correspond to the  dispersionless
generalized dKP hierarchy briefly discussed in Sec.2. This remark
closes the study of the orbit defined by the symmetry action.

\section{The process of gluing.}

An elementary way of gluing together two algebras of
polynomials, one in the variable p and the other in
the variable q, is to add the relation
\begin{equation*}
pq=ap+bq+c
\end{equation*}
which allows to write the product of a polynomial
in p by a polynomial in q as sum of two polynomials
of the same variables. The new associative algebra is known
as the quotient of the algebra of polynomials in two
variables p and q with respect to the ideal generated
by the polynomial $(-pq+ap+bq+c)$.

The leading idea of this section is to apply this procedure
to two copies of the algebra of Faa' di Bruno polynomials
encountered in Sec.3 and slightly generalized in Sec.5.
In practice this means that we complete the tables of
multiplication
\begin{gather}
\label{gluing}
\begin{aligned}
p_{j}p_{k} &=& p_{j+k}+\sum_{l=0}^{j}H_{l}^{j}p_{k-l}+
\sum_{l=0}^{k}H_{l}^{k}p_{j-l}+D_{jk},  \\
q_{j}q_{k} &=& q_{j+k}+\sum_{l=0}^{j}\widetilde{H}_{l}^{j}q_{k-l}+
\sum_{l=0}^{k}\widetilde{H}_{l}^{k}q_{j-l}+\widetilde{D}_{jk}
\end{aligned}
\end{gather}
by adding the relation
\begin{equation}
\label{gluing2} p_{1}q_{1}=ap_{1}+bq_{1}+c  ,
\end{equation}
and consistently computing the products
$p_{j}q_{k}$. The aim is to show that the structure
constants of the enlarged table of multiplication
admit interesting coisotropic deformations reproducing
the universal Whitham equations of genus zero, obtained by
Krichever by his technique of meromorphic functions
on the Riemann sphere with punctures \cite{Kri3}. The process of gluing
is thus the algebraic implementation of the pasting of
meromorphic functions on the Riemann sphere (or of
meromorphic differentials on Riemann surfaces)
characteristic of the algebro-geometric approach
of Krichever. What for him is the introduction of
a new pole, for us is simply the gluing of an additional
copy of the same algebra.

To give an idea of the potentialities of the new procedure,
let us consider the fragment of the above
table of multiplication consisting of the three simplest
equations
\begin{gather}
\label{partial table}
\begin{aligned}
p_{1}p_{1} &= p_{2}-(vp_{1}+u), \\
q_{1}q_{1} &= q_{2}-(\tilde{v}p_{1}+\tilde{u}),  \\
p_{1}q_{1} &= ap_{1}+bq_{1}+c  ,
\end{aligned}
\end{gather}
where new names have been used for the structure constants to
simplify the notation. According to the scheme of coisotropic
deformations, we introduce two sets of space coordinates $x_{j}$
and $y_{j}$ and their conjugate momenta $p_{j}$ and $q_{j}$, and
we transform the partial table of multiplication~(\ref{partial
table}) into the definition of three Hamiltonians
\begin{gather}
\begin{aligned}
f &= -p_{2}+p_{1}^{2}+ vp_{1}+u, \\
\tilde{f} &= -q_{2}+q_{1}^{2}+\tilde{v}p_{1}+\tilde{u}, \\
g &= -p_{1}q_{1}+ap_{1}+bq_{1}+c
\end{aligned}
\end{gather}
on the symplectic manifold $R^{8}$ endowed with
the classical Poisson bracket for canonical coordinates
$(x_{1},x_{2},y_{1},y_{2},p_{1},p_{2},q_{1},q_{2})$.
Finally we demand that the ideal $J$ generated by
these Hamiltonians be closed with respect to the Poisson
bracket, and we obtain three sets of equations. From the study of
$\{ f,g \}$ we obtain the equations
\begin{align*}
v_{y_{1}}+2a_{x{1}} =0,  \\
a_{x_{2}}-2(ab+c)_{x_{1}}-(1/4v^2+u)_{y_{1}}=0, \\
b_{x_{2}}-(bv+b^2-u)_{x_{1}}=0, \\
c_{x_{2}}-(cv)_{x_{1}}-2cb_{x_{1}}+au_{x_{1}}+bu_{y_{1}}=0, \\
\end{align*}
while the study of $\{ \tilde{f},g  \}$ leads to
\begin{align*}
2b_{y_{1}}+\tilde{v}_{x_{1}}=0,  \\
a_{y_{2}}+(a\tilde{v}-a^{2}-\tilde{u})_{y_{1}}=0,  \\
b_{y_{2}}-2(ab+c)_{y_{1}}-(\tilde{u}+1/4\tilde{v}^{2})_{x_{1}}=0, \\
c_{y_{2}}-(c\tilde{v})_{y_{1}}-2ca_{y_{1}}+a\tilde{u}_{x_{1}}+
b\tilde{u}_{y_{1}}=0,
\end{align*}
and the study of $\{ f,\tilde{f} \}$ gives
\begin{align*}
2av_{y_{1}}-2a\tilde{v}_{x_{1}}+\tilde{v}v_{y_{1}}-v_{y_{2}}-
2\tilde{u}_{x_{1}}=0, \\
2bv_{y_{1}}-2b\tilde{v}_{x_{1}}-v\tilde{v}_{x_{1}}-
\tilde{v}_{x_{2}}-2\tilde{u}_{y_{1}}=0, \\
2cv_{y_{1}}-2c\tilde{v}_{x_{1}}+\tilde{v}u_{y_{1}}-u_{y_{2}}-
v\tilde{u}_{x_{1}}+\tilde{u}_{x_{2}}-u_{y_{2}}=0.
\end{align*}
If these equations are satisfied then
\begin{align*}
\{f,g \} &= \left(v+2b \right)_{x_{1}} g, \\
\{\tilde{f},g \} &= \left(\tilde{v}+ 2 a \right)_{y_{1}} g,\\
\{f,\tilde{f} \} &= 2 \left(\tilde{v}_{x_{1}} - v_{y_{1}} \right)
g
\end{align*}
 These equations admit several interesting reductions. For
instance, by setting $v=0, \tilde{u}=0,a=0$ the first system
becomes
\begin{displaymath}
u_{y_{1}}+2c_{x_{1}}=0, \qquad b_{x_{2}}-(b^2+u) _{x_{1}}=0,
\qquad c_{x_{2}}-2(bc)_{x_{1}}=0
\end{displaymath}
which is the simplest example of generalized Benney's system
introduced in~\cite{Kup2,Kri3,Za3}. At the same time the second
system becomes the modified Benney's system~\cite{Kup2}
\begin{displaymath}
\tilde{v}_{x_{1}}+2b_{y_{1}}=0,  \qquad
b_{y_{2}}-1/4(\tilde{v}^{2})_{x_{1}}-2c_{y_{1}}=0,  \qquad
c_{y_{2}}-(c\tilde{v})_{y_{1}}=0,
\end{displaymath}
and the third system becomes the Miura type
transformation
\begin{displaymath}
\tilde{v}_{x_{2}}+2(u+b^{2})_{y_{1}}=0,  \qquad
u_{y_{2}}+2(c\tilde{v})_{x_{1}}=0
\end{displaymath}
between them~\cite{Kup2}. If instead one sets $u=0,\tilde{u}=0,
c=0$ from the first two systems of coisotropy conditions one
obtains
\begin{align*}
a_{x_{2}}-2(ab)_{x_{1}}-1/4(v^{2})_{y_{1}}=0,  \\
b_{x_{2}}-(b^{2}+bv)_{x_{1}}=0,  \\
v_{y_{1}}+2a_{x_{1}}=0,  \\
a_{y_{2}}-(a^{2}+a\tilde{v})_{y_{1}}=0,  \\
b_{y_{2}}-2(ab)_{y_{1}}-1/4(\tilde{v}^{2})_{x_{1}}=0,  \\
\tilde{v}_{x_{1}}+2b_{y_{1}}=0
\end{align*}
which is, in fact, the dispersionless limit of the
equations for the components of the wave function
of the Dawey-Stewartson equation discussed in \cite{K7}.

It is remarkable that the new associative algebra, defined by the
process of gluing, is isomorphic to the algebra of meromorphic
functions on the Riemann sphere with two punctures.Indeed
resolving the coupling relation with respect to $q_{1}$, one gets
\begin{equation*}
q_{1}=\frac{ab+c}{p_{1}-b} +a .
\end{equation*}
Hence any polynomial in $q_{1}$ becomes a rational function in
$p_{1}$. So the algebra~(\ref{gluing}), (\ref{gluing2}) may be
identified with the algebra of rational functions with poles in
$b$ and in the point at infinity. This algebra is just the algebra
of meromorphic functions on the Riemann sphere with two punctures
used by Krichever \cite{Kri3}. The general case of n puntures is
equivalent to the gluing of n copies of algebras of the initial
type.

To give an example of equations coming from the
gluing of $n$ copies of the algebra of Faa' di Bruno
polynomials, let us consider the fragment of the table
of multiplication consisting simply of the $1/2n(n-1)$
relations
\begin{equation*}
p_{\alpha}p_{\beta}=H_{\alpha}^{\beta}p_{\alpha}+
H_{\beta}^{\alpha}p_{\beta}  \qquad  \alpha\neq{\beta}
\end{equation*}
which serve to glue together the n copies of the algebra.
In this example the indices $\alpha,\beta=1,2,\ldots,n$
identify the copies of the algebra, and the symbol $p_{\alpha}$
stands for $p_{\alpha,1}$.

If n is greater than two, the coefficients $H_{\beta}^{\alpha}$
must verify an appropriate set of associativity conditions.
In Appendix B it is shown that the part of the associativity
conditions pertaining to the coefficients $H_{\beta}^{\alpha}$
is
\begin{equation}
\label{associativity3} H_{\alpha}^{\beta}H_{\alpha}^{\gamma}-
H_{\gamma}^{\beta}H_{\alpha}^{\gamma}-
H_{\beta}^{\gamma}H_{\alpha}^{\beta}=0  ,
\end{equation}
where the indices $\alpha,\beta,\gamma$ take distinct values.
In the same Appendix it is also shown that the coisotropy
conditions for the Hamiltonians
\begin{equation*}
g_{\alpha\beta}=-p_{\alpha}p_{\beta}+
H_{\alpha}^{\beta}p_{\alpha}+H_{\beta}^{\alpha}p_{\beta} ,
\end{equation*}
$\alpha \neq  \beta $, contains a subset of conditions
pertaining only to the coupling coefficients
$H_{\beta}^{\alpha}$, and that this subset simplifies
into the by now familiar form
\begin{equation*}
\frac{\partial {H_{\gamma}^{\alpha}}}{\partial {x_{\beta}}}-
\frac{\partial {H_{\gamma}^{\beta}}}{\partial {x_{\alpha}}}=0
\qquad \alpha \neq{\beta}\neq{\gamma}\neq {\alpha}
\end{equation*}
owing to the previous associativity conditions. The
equations just written describe the coisotropic deformations
of the coupling coefficients $H_{\beta}^{\alpha}$ alone.

As in the case of the dispersionless KP hierarchy, one may
treat these equations in two alternative ways. If one solves
first the associativity conditions by setting, for instance,
\begin{equation*}
H_{\beta}^{\alpha}=\frac{u_{\alpha}}{v_{\alpha}-v_{\beta}}
\qquad  \alpha\neq{\beta}  ,
\end{equation*}
where $u_{\alpha}$ and $v_{\alpha}$ are arbitrary functions
of the coordinates $x_{\alpha}$, one obtains from the
coisotropy conditions the final system
\begin{equation*}
\frac{\partial}{\partial {x_{\beta}}}
(\frac{u_{\alpha}}{v_{\alpha}-v_{\gamma}})-
\frac{\partial}{\partial {x_{\alpha}}}
(\frac{u_{\beta}}{v_{\beta}-v_{\gamma}})=0  .
\end{equation*}
If instead one chooses the second strategy, and solves first
the coisotropy conditions by setting
\begin{equation*}
H_{\beta}^{\alpha}=\frac{\partial {F_{\beta}}}{\partial {x_{\alpha}}}
\qquad   \alpha \neq {\beta}  ,
\end{equation*}
one obtains from the associativity conditions the
dispersionless Darboux system
\begin{equation*}
\frac{\partial {F_{\alpha}}}{\partial {x_{\beta}}}
\frac{\partial {F_{\alpha}}}{\partial {x_{\gamma}}}-
\frac{\partial {F_{\alpha}}}{\partial {x_{\gamma}}}
\frac{\partial {F_{\gamma}}}{\partial {x_{\beta}}}-
\frac{\partial {F_{\alpha}}}{\partial {x_{\beta}}}
\frac{\partial {F_{\beta}}}{\partial {x_{\gamma}}}=0
\end{equation*}
introduced in \cite{K3}. It represents the dispersionless limit
of the well-known Darboux system describing conjugate
nets of curves in $R^{n}$ ( see e.g. \cite{Dar}).

The second example is the n-component (2+1)-dimensional Benney
system introduced by Zakharov \cite{Za3}. It can be recovered by gluing
one copy of the algebra associated with the dispersionless
KP hierarchy to $(n+1)$ copies of the algebras associated with
the dispersionless mKP hierarchy. The Benney systems can
then obtained as coisotropy conditions of the ideal $J$ generated by the
first Hamiltonian of the dispersionless KP hierarchy
\begin{equation*}
f = -p_{2}+p_{1}^2-u  ,
\end{equation*}
by the $n$ functions
\begin{equation*}
k_{\alpha} = p_{1} q_{\alpha} -a_{\alpha} q_{\alpha}- \nu_{\alpha} \qquad \alpha=1,\ldots,n  \\
\end{equation*}
which serve to glue the algebra of the dispersionless KP
hierarchy to $n$ copies of the algebra of the
dispersionless mKP hierarchy, and by the function
\begin{equation*}
g=q_{n+1}-\sum_{\alpha=1}^{n}q_{\alpha} ,
\end{equation*}
which serves to glue the $(n+1)-th$ copy of this algebra to the
previous ones. To apply the technique of coisotropic deformations
we have to simultaneously introduce new deformation parameters.
Let us denote by $(x_{1},x_{2},y_{\alpha},\tau )$ the coordinates
conjugate to the momenta $(p_{1},p_{2},q_{\alpha},q_{n+1})$
respectively, and let us work out explicitly the coisotropy
conditions of the ideal $J$. One may easily find that the
condition $\{f,k_{\alpha}\}=0$ on $\Gamma$ gives
\begin{align}\label{uy}
u_{y_{\alpha}}+2\nu_{\alpha x_{1}}&=0, \\
\label{ux1} u_{x_{1}}+(a_{\alpha}^{2})_{x_{1}}-a_{\alpha
x_{2}}&=0,
\qquad \alpha=1,\ldots,n \\
\label{nualpha} \nu_{\alpha
x_{2}}+a_{\alpha}u_{y_{\alpha}}-2\nu_{\alpha}a_{\alpha x_{1}}&=0
\end{align}
Similarly the condition $\{f,g\}=0$ on $\Gamma$ implies
\begin{equation}\label{utau}
u_{\tau}-\sum_{\alpha=1}^{n}u_{y_{\alpha}}=0
\end{equation}
while the conditions $\{k_{\alpha},g\}=0 $ on $\Gamma$ give
\begin{equation*}
a_{\alpha \tau }-\sum_{\beta=1}^{n} \frac{\partial {a}_{\alpha} }{\partial {y}_{\beta} }=0
\end{equation*}
and
\begin{equation*}
\nu_{\alpha \tau }-\sum_{\beta=1}^{n}\frac{\partial {\nu}_{\alpha} }{ \partial {y}_{\beta}}=0  .
\end{equation*}
From equations (\ref{uy}) and (\ref{utau}) one gets
\begin{equation*}
u_{\tau}+2(\sum_{\alpha=1}^{n}\nu_{\alpha})_{x_{1}}=0.
\end{equation*}
Then from equations (\ref{uy}) and equations (\ref{nualpha}) one obtains
\begin{equation*}
\nu_{{\alpha x_{2}}_2}-2( \nu_{\alpha} a_{\alpha})_{x_{1}} =0 .
\end{equation*}
The last two equations together with (\ref{ux1}) are just the
(2+1)-dimensional n-component Benney's system \cite{Za3}.

If one wants to glue $n$ copies of the full
algebras
\begin{equation*}
p_{k}p_{j} = p_{k+j}+\sum_{l=1}^{j}H_{l}^{k}p_{j-l}+\sum_{l=1}^{k}H_{l}^{j}p_{k-l}+D_{jk} ,
\end{equation*}
by means of the gluing relations
\begin{equation*}
p_{\alpha,i} p_{\beta,k}=\sum_{\gamma=1}^{n}\sum_{l\geq 1 }
C_{\alpha \beta ik}^{\gamma l}p_{\gamma l} + C_{\alpha \beta ik}^{0} p_{0} ,
\end{equation*}
where $p_{0}$ is the unit element, $\alpha,\beta=1,\ldots,n$ and
$i,k,l=1,2,\ldots $, one is obliged to consider the full system of
structure constants
\begin{equation*}
C_{\alpha \beta ik}^{\gamma l}=\delta_{\gamma \alpha} \delta_{\gamma \beta} \delta_{i+k}^{l}+
\delta_{\gamma \alpha} H_{\alpha,i-l}^{\beta,k}+
\delta_{\gamma \beta} H_{\beta,k-l}^{\alpha,i}  .
\end{equation*}
Coisotropic deformations for this algebra can be constructed
according to our general scheme, but they will be studied elsewhere.

\section{Coisotropic and Lagrangian submanifolds.}

Our purpose in this section is to stress an important
geometrical difference between the usual approach to
integrable dispersionless equations, based on the study
of compatibility conditions of a system of Hamilton-
Jacobi equations, and the present approach of coisotropic
deformations.

A characteristic trait of the usual approach is to assume
$p_{j}=\frac{\partial {S}}{\partial {x_{j}}}$ since the beginning,
and therefore one is unwittingly confined to a Lagrangian
submanifold inside the symplectic manifold $M^{2n}$. As a
consequence one looses the canonical symplectic 2-form
\begin{equation*}
\omega=\sum_{i=1}^{n}dp_{i}\wedge{dx_{i}}
\end{equation*}
which vanishes on the Lagrangian submanifold. In the present
approach, instead, the main role is given the coisotropic
submanifold $\Gamma$ (for the definition of coisotropic
submanifolds see e.g. \cite{Wein1}-\cite{Ber}). It is interesting
to see some examples of them. Any solution of the integrable
hierarchies discussed in the paper provides us with a coisotropic
submanifold. A simple example corresponds to the following
solution of the dKP equation (\cite{Kod1})
\begin{eqnarray*}
u_{0} &=& -2/3\frac{x_{1}}{(1+2x_{3})}
+4/9\frac{x_{2}^{2}}{(1+2x_{3})^{2}},  \\
v_{0} &=& -2/3\frac{x_{1}x_{2}}{(1+2x_{3})^{2}}
+4/9\frac{x_{2}^{3}}{(1+2x_{3})^{3}}.\\
\end{eqnarray*}
The associated coisotropic submanifold of dimension 4 in the six
dimensional symplectic space $R^{6}$, with coordinates
$(x_{1},x_{2},x_{3},p_{1},p_{2},p_{3})$ is defined by the equations
\begin{eqnarray*}
f_{11} &=& -p_{1}^{2}+p_{2} +2/3\frac{x_{1}}{(1+2x_{3})}
-4/9\frac{x_{2}^{2}}{(1+2x_{3})^{2}} =0, \\
f_{12} &=& -p_{1}p_{2}+p_{3}-p_{1}( -1/3\frac{x_{1}}{(1+2x_{3})}
+2/9\frac{x_{2}^{2}}{(1+2x_{3})^{2}})
-2/3\frac{x_{1}x_{2}}{(1+2x_{3})^{2}}\\
&+&4/9\frac{x_{2}^{3}}{(1+2x_{3})^{3}}  =0 .\\
\end{eqnarray*}

If instead one takes the zero-set of the first three Hamiltonians
$(f_{11},f_{12},f_{13})$ of the dKP hierarchy, and one considers
the common solution $u_{0}(x_{1},x_{2},x_{3},x_{4})$ of the dKP
and first higher dKP equations, one gets a five dimensional
coisotropic submanifold in $R^{8}$. Continuing in this process,
one obtains an infinite tower on coisotropic submanifolds
associated with dKP hierarchy. In general they have dimension
$(n+1)$ in a symplectic space of dimension $2n$. In this sense
they are minimal since they are the most close to Lagrangian
submanifolds which have dimension $n$. Furthermore, the
restriction of the symplectic 2-form $\omega$ to $\Gamma$ does not
vanish, but is a presymplectic 2-form. As is well-known, its
kernel is spanned by the (n-1) Hamiltonian vector fields
associated with the functions $f_{jk}$ defining the coisotropic
submanifold. These vector fields define the so-called
characteristic foliation of $\Gamma$, whose space of leaves is
called the reduced phase space $\Gamma_{red}$ (see e.g.
\cite{Wein1}-\cite{Ber}). In our case the restriction of $\omega$
to $\Gamma$ has rank two and hence
\begin{equation}\label{restrizione}
\omega_{\Gamma}= d \textit{L}\wedge d\textit{M} .
\end{equation}
The canonical variables \textit{L} and \textit{M} play an important
role in the theory of dispersionless hierarchies.

Various Lagrangian submanifolds are obtained by setting a
constraint on these variables. An obvious example is provided by
the constraint
\begin{equation}
\textit{L}=z=const .
\end{equation}
As usual the Lagrangianity implies the existence of a generating function $S(z,x)$ such that (see e.g. \cite{Ar1,Ar2,Wein2})
\begin{equation}
p_{j}(x)=\frac{\partial {S}(z,x)}{\partial {x_{j}}} \quad
\textup{for} \qquad z=const.
\end{equation}
One also readily concludes that the restriction of \textit{M} to the Lagragian submanifold
is given by
\begin{equation}
\textit{M}|_{z}=M(z)=\frac{\partial {S}(z,x)}{\partial {z}} .
\end{equation}
With this identification  the formula~(\ref{restrizione})
coincides with that obtained in \cite{TT1,Kri3}. Thus one may say
that the foliation of the coisotropic submanifold $\Gamma$ by
Lagrangian submanifolds parametrized by $z$ corresponds to the
``algebraic orbits" of the Whitham's equations of genus zero
considered in \cite{Kri3}. A more general class of Lagrangian
submanifolds are provided by the constraint
\begin{equation}
g(\textit{L},\textit{M})=0
\end{equation}
where g is an arbitrary function.

Finally we would like to note that the nonlinear $\bar{\partial} $ equation,
which is the basic equation for the quasiclassical $\bar{\partial} $ method (\cite{K1,K2,K3}),
also has a simple geometric meaning within the present approach, at least if one
considers the complexified version of the scheme, where all the variables are complex.
In this case the rank of $\omega|_{\Gamma}$ is equal to four ( complex two), and
\begin{equation}
\omega |_{\Gamma}=d \textit{L} \wedge d \textit{M}+ d \bar{\textit{L}}  \wedge {d\tilde{\textit{M}}}
\end{equation}
where the bar denotes, as usual, the complex conjugation. In this
case the class of Lagrangian submanifolds is defined by the
constraint
\begin{equation}\label{W}
W(\textit{L},\bar{\textit{L}},\textit{M},\tilde{\textit{M}})=0
\end{equation}
where W is an arbitrary complex function. Using the parametrization
$\textit{L}=z,\bar{\textit{L}}=\bar{z}$ and the formulae $\textit{M}=\frac{\partial {S}}{\partial {z}},
\widetilde{\textit{M}}=\frac{\partial {S}}{\partial \bar{z}}$, one writes equation (\ref{W}) as
\begin{equation}
W(z,\bar{z},\frac{\partial {S}}{\partial {z}},\frac{\partial {S}}{\partial \bar{z}})=0
\end{equation}
which is exactly the nonlinear $\bar{\partial} $ equation used in
the papers \cite{K1,K2,K3,K4,K5}. So many of the known methods to
solve dispersionless integrable equations deal, actually, with
different classes of Lagrangian submanifolds contained inside the
coisotropic submanifold $\Gamma$.

\section{Final remarks.}

The aim of this paper was to point out a new phenomenon (the
appearance of structure constants inside the theory of
dispersionless integrable hierarchies), without pretention of
completeness or systematicity. The paper is, indeed, a first
exploration of a territory which remains largely unknown. A few
other directions of explorations are known. Some are routine work,
consisting in encompassing different system of structure constants
and therefore different classes of integrable hierarchies, such as
, for instance, the dispersionless Harry Dym hierarchy. Some have
the theoretical aim of probing more deeply the basis of the
Hirota's bilinear formulation, by understanding for what kind of
associative algebras the associativity conditions allow to reduce
the coisotropy conditions to the form of a system of conservation
laws. In the same vain, the other interesting question is to
understand systematically if the concept of coisotropy has nothing
to do with a possible Hamiltonian (or bihamiltonian) structure of
the integrable hierarchy defined by the central system on the
structure constants. A last exciting direction, finally, points
towards more general type of algebras. We know that to encompass,
for instance, the dispersionless Veselov-Novikov hierarchy in the
present scheme, one has to abandon the associative commutative
algebra with unit and to consider Jordan's triple systems. So of
the scheme presented in this paper one should take mainly the
spirit rather than the form, and to consider  it as a possible
point of departure for new interesting investigations in a field
which has not yet exhausted is source of surprises, despite
intensive investigations during the last twenty years or more.

\section{Appendix A.}

Here we will derive equations~(\ref{associativity2}),
(\ref{symmetry relations}) and (\ref{exactness}) for the
generalised dKP hierarchy.

Thus we consider an algebra with the structure constants of the
form
\begin{equation}
\label{structure constants2} C_{jk}^{l} = \delta_{j+k}^{l} +
H_{j-l}^{k} + H_{k-l}^{j} + \delta_{0}^{l} D_{jk}
\end{equation}
where $H_{0}^{l} \neq 0$, $H_{k}^{j} = 0$ $k \leq -1$.

 For such the structure constants the
associativity condition takes the form
\begin{gather}
\begin{aligned}\label{prima}
R_{ikmn}
&+H_{-n}^{m}D_{ik}-H_{-n}^{i}D_{mk}+\delta_{m}^{n}D_{ik}-\delta_{i}^{n}D_{mk}\\
&+\delta_{0}^{n}\left( D_{i+k,m}-D_{i,m+k}+D_{ik}D_{0m}-D_{mk}D_{0i} \right ) \\
&+\delta_{0}^{n}\left
(\sum_{l=0}^{k}H_{k-l}^{i}D_{lm}+\sum_{l=0}^{i}H_{i-l}^{k}D_{lm}-\sum_{l=0}^{k}H_{k-l}^{m}D_{li}
-\sum_{l=0}^{m}H_{m-l}^{k}D_{li} \right) =0
\end{aligned}
\end{gather}
where $R_{ikmn}$ is given by l.h.s of
equation~(\ref{associativity0}) with $H_{0}^{i} \neq 0$ and
substitution $\sum_{l=1} \rightarrow \sum_{0}$.
Equation~(\ref{prima}) in the case when all indices
$i$,$k$,$m$,$n$ are distinct and different from zero is of the
form
\begin{equation}
R_{ikmn} = 0.
\end{equation}
 These equations are easily seen to be equivalent to the
system
\begin{equation}\label{seconda}
\sum_{m+k=p} [ H,H ]_{ikm} = 0 ,
\end{equation}
where
\begin{equation*}
\left [ H,H \right ]_{ikm}:= H_{m}^{i+k}-H_{m+k}^{i}-H_{i+m}^{k}
+\sum_{l=0}^{i}H_{i-l}^{k}H_{m}^{l}+\sum_{l=0}^{k}H_{k-l}^{i}H_{m}^{l}
-\sum_{l=0}^{m}H_{m-l}^{k}H_{l}^{i}.
\end{equation*}
Note that these equations do not contain $H_{0}^{i}$.

Equations (\ref{prima}) with $n=m\neq{0}$ and all other indices
distinct are equivalent to
\begin{equation}\label{terza}
D_{ik}+H_{0}^{i+k}-H_{0}^{i}H_{0}^{k}+\sum_{l=1}^{k}H_{k-l}^{i}H_{0}^{l}
+\sum_{l=1}^{i}H_{i-l}^{i}H_{0}^{l} =0 .
\end{equation}
At $n=i\neq{0}$ one gets equations which are equivalent to these ones.

Finally at $n=0$ equation (\ref{prima}) is reduced to
\begin{gather*}
\begin{aligned}
&D_{i+k,m}-D_{i,m+k}+\sum_{l=0}^{k}H_{k-l}^{i}D_{lm} +\sum_{l=0}^{i}H_{i-l}^{k}D_{lm} \\
&-\sum_{l=0}^{k}H_{k-l}^{m}D_{li}-\sum_{l=0}^{m}H_{m-l}^{k}D_{li}
=0 .
\end{aligned}
\end{gather*}
It is a straightforward but cumbersome check that these equations are verified in virtue of
the previous ones. So the associativity conditions for the structure constants
(\ref{newstructureconstants}) are given by equations (\ref{seconda}) and (\ref{terza}).

The coisotropy conditions  $ \{ f_{ik},f_{ln} \} =0$ on $\Gamma$
for the Hamiltonians defined by the above structure constants are
equivalent to the system
\begin{gather}
\begin{aligned}
\label{quarta}
T_{ikln}+\delta_{0}^{m}Q_{ikln}-\delta_{n}^{m}\frac{\partial
C_{ik}^{0}}{\partial x_{l}} -\delta_{l}^{m}\frac{\partial
C_{ik}^{0}}{\partial x_{n}}+\delta_{k}^{m}\frac{\partial
C_{ln}^{0}}{\partial x_{i}}+\delta_{i}^{m}\frac{\partial
C_{ln}^{0}}{\partial x_{k}}=0  ,
\end{aligned}
\end{gather}
where
\begin{equation*}
C_{ik}^{0}=H_{k}^{i}+H_{i}^{k}+D_{ik}
\end{equation*}
and
\begin{gather*}
\begin{aligned}
T_{ikln} &=\sum_{s=1}\left(\delta_{l+n}^{s}
+H_{n-s}^{l}+H_{l-s}^{n}\right)\left(\frac{\partial
H_{k-m}^{i}}{\partial x_{s}}+ \frac{\partial H_{i-m}^{k}}{\partial
x_{s}}\right) \\
&- \sum_{s=1}\left(\delta_{i+k}^{s}
+H_{k-s}^{i}+H_{i-s}^{k}\right)\left(\frac{\partial
H_{n-m}^{l}}{\partial x_{s}}+
\frac{\partial H_{l-m}^{n}}{\partial x_{s}}\right)  \\
&-\sum_{s=1}\left(\delta_{s+n}^{m}
+H_{n-m}^{s}+H_{s-m}^{n}\right)\left(\frac{\partial
H_{k-s}^{i}}{\partial x_{l}}+ \frac{\partial H_{i-s}^{k}}{\partial
x_{l}}\right) \\
&- \sum_{s=1}\left(\delta_{l+s}^{m}
+H_{l-m}^{s}+H_{s-m}^{l}\right)\left(\frac{\partial
H_{k-s}^{i}}{\partial x_{n}}+
\frac{\partial H_{i-s}^{k}}{\partial x_{n}}\right) \\
&+\sum_{s=1}\left(\delta_{k+s}^{m}
+H_{s-m}^{k}+H_{k-m}^{s}\right)\left(\frac{\partial
H_{n-s}^{l}}{\partial x_{i}}+ \frac{\partial H_{l-s}^{k}}{\partial
x_{i}}\right) \\
&+ \sum_{s=1}\left(\delta_{i+s}^{m}
+H_{s-m}^{i}+H_{i-m}^{s}\right)\left(\frac{\partial
H_{n-s}^{l}}{\partial x_{k}}+ \frac{\partial H_{l-s}^{n}}{\partial
x_{k}}\right)
\end{aligned}
\end{gather*}
and
\begin{gather*}
\begin{aligned}
Q_{ikln}&=\frac{\partial D_{ik}}{\partial x_{l+n}}-\frac{\partial
D_{ln}}{\partial x_{i+k}}+
\sum_{s=0}\lbrack(H_{n-s}^{l}+H_{l-s}^{n})\frac{\partial
D_{ik}}{\partial x_{s}}
-(H_{k-s}^{i}+H_{i-s}^{k})\frac{\partial D_{ln}}{\partial x_{s}} \\
&-D_{sn}(\frac{\partial H_{k-s}^{i}}{\partial
x_{l}}+\frac{\partial H_{i-s}^{k}}{\partial x_{l}})-
D_{sl}(\frac{\partial H_{k-s}^{i}}{\partial x_{n}}+\frac{\partial H_{i-s}^{k}}{\partial x_{n}}) \\
&+D_{sk}(\frac{\partial H_{n-s}^{l}}{\partial
x_{i}}+\frac{\partial H_{l-s}^{n}}{\partial x_{i}})+
D_{si}(\frac{\partial H_{n-s}^{l}}{\partial x_{k}}+\frac{\partial
H_{l-s}^{n}}{\partial x_{k}})\rbrack .
\end{aligned}
\end{gather*}
For arbitrary indices $i,k,l,n$ and for $m>i,k,l,n$ ;$m>n+i$ ; $m>l+i$ ; $m>l+k$ ;$m\leq{n+k}$
equations (\ref{quarta}) are reduced to
\begin{equation*}
-\frac{\partial H_{k+n-m}^{i}}{\partial x_{l}}+\frac{\partial H_{k+n-m}^{l}}{\partial x_{i}}=0 .
\end{equation*}. Since $k+n-m\geq{0}$ one has therefore
\begin{equation}\label{sesta}
\frac{\partial H_{p}^{i}}{\partial x_{l}}=\frac{\partial H_{p}^{l}}{\partial x_{i}}
\end{equation}
for $p\geq{0}$  and $ i,l\geq{1}$.

Using this condition, one can show that $T_{iklnm}$ can be
represented as
\begin{equation*}
T_{iklnm}=\frac{\partial \left [ H,H \right ]_{l,n,k-m}}{\partial x_{i}}+
\frac{\partial \left [ H,H \right ]_{l,n,i-m}}{\partial x_{k}}-
\frac{\partial \left [ H,H \right ]_{i,k,l-m}}{\partial x_{n}}-
\frac{\partial \left [ H,H \right ]_{i,k,n-m}}{\partial x_{l}}  .
\end{equation*}
So equations~(\ref{quarta}) take the form
\begin{gather}
\begin{aligned}
\label{quinta}
&\frac{\partial}{\partial x_{i}}\left( \left [ H,H \right ]_{l,n,k-m}+
\delta_{k}^{m}C_{ln}^{0}\right)+
\frac{\partial}{\partial x_{k}}\left( \left [ H,H \right ]_{l,n,i-m}+
\delta_{i}^{m}C_{ln}^{0}\right)  \\
&-\frac{\partial}{\partial x_{n}}\left( \left [ H,H \right ]_{i,k,l-m}+
\delta_{l}^{m}C_{ik}^{0}\right)-
\frac{\partial}{\partial x_{l}}\left( \left [ H,H \right ]_{i,k,n-m}+
\delta_{n}^{m}C_{ik}^{0}\right)+\delta_{0}^{m}Q_{ikln}=0  .
\end{aligned}
\end{gather}
For arbitrary indices $i,k,l,n$ and for $m<i,k,l,n$ ; $m\neq{0}$ these equations
are reduced to
\begin{equation*}
\frac{\partial \left [ H,H \right ]_{l,n,k-m}}{\partial x_{i}}+
\frac{\partial \left [ H,H \right ]_{l,n,i-m}}{\partial x_{k}}-
\frac{\partial \left [ H,H \right ]_{i,k,l-m}}{\partial x_{n}}-
\frac{\partial \left [ H,H \right ]_{i,k,n-m}}{\partial x_{l}}=0  .
\end{equation*}
These equations are satisfied in virtue of the associativity conditions.

For arbitrary indices $i,k,l,n$ and for $m\neq{k}$ ; $m<i,l,n$ ; $m\neq{0}$ equation
(\ref{quinta}) is
\begin{equation*}
\frac{\partial}{\partial x_{i}}\left(\left [ H,H \right ]_{l,n,0}+C_{ln}^{0}\right)=0.
\end{equation*}
It is satisfied if
\begin{equation*}
\left [ H,H \right ]_{l,n,0}+C_{ln}^{0}=0  .
\end{equation*}
Since
\begin{gather*}
\begin{aligned}
\left [ H,H \right ]_{l,n,0} &= H_{0}^{l+n}-H_{n}^{l}-H_{l}^{n}-H_{0}^{l}H_{0}^{n} \\
&+\sum_{s=0}^{l}H_{l-s}^{n}H_{0}^{s}+\sum_{s=0}^{n}H_{n-s}^{l}H_{0}^{s}
\end{aligned}
\end{gather*}
the above equations just follow from the associativity conditions.
In the same manner one can show that in the cases $m\neq{0}$ and
$m=i$ or $m=l$ or $m=n$ equations (\ref{quinta}) are satisfied
owing to the associativity conditions.

At $m=0$ and arbitrary $i,k,l,n\neq{0}$ equations (\ref{quinta})
take the form
\begin{gather*}
\begin{aligned}
\frac{\partial \left [ H,H \right ]_{l,n,k}}{\partial x_{i}}+
\frac{\partial \left [ H,H \right ]_{l,n,i}}{\partial x_{k}} -
\frac{\partial \left [ H,H \right ]_{i,k,l}}{\partial x_{n}}
-\frac{\partial \left [ H,H \right ]_{i,k,n}}{\partial x_{l}}
+Q_{ikln}=0
\end{aligned}
\end{gather*}and, consequently, they reduce to
\begin{equation*}
Q_{ikln}=0  .
\end{equation*}
It is a direct but quite cumbersome check that these equations also are satisfied in virtue of
equations (\ref{seconda}),(\ref{terza}), and (\ref{sesta}).

Thus, the coisotropic deformations of the generalized dKP algebra
defined by the structure constants (\ref{newstructureconstants})
are given by the associativity conditions (\ref{seconda}),
(\ref{terza}), and the exactness conditions (\ref{sesta}). The dKP
case corresponds to the particular choice of $H_{0}^{i} = 0$ and
$D_{jk} = 0$.

\section{Appendix B.}

For the algebra given by
\begin{equation}
\label{1B} p_{\alpha}p_{\beta}= H_{\alpha}^{\beta}
p_{\alpha}+H_{\beta}^{\alpha} p_{\beta}+ C_{\alpha \beta}^{0}
\quad{}  \alpha \neq \beta
\end{equation}
the associativity conditions give rise to the system
\begin{equation}
\label{2B} C_{\alpha
\beta}^{0}-H_{\gamma}^{\alpha}H_{\gamma}^{\beta}+
H_{\gamma}^{\alpha}H_{\alpha}^{\beta}+H_{\gamma}^{\beta}H_{\beta}^{\alpha}=0
\qquad
 \alpha\neq{\beta}\neq{\gamma}\neq{\alpha}
\end{equation}
and
\begin{equation}
\label{3B}
H_{\alpha}^{\beta}C_{\alpha\gamma}^{0}+H_{\beta}^{\alpha}C_{\beta\gamma}^{0}-
H_{\beta}^{\gamma}C_{\alpha\gamma}^{0}-H_{\gamma}^{\beta}C_{\alpha\gamma}^{0}=0 .
\end{equation}
Equations (\ref{2B}) imply $1/2n(n-1)(n-2)$ equations
\begin{gather}
\begin{aligned}
\label{4B}
&H_{\gamma}^{\alpha}H_{\gamma}^{\beta}-
H_{\gamma}^{\alpha}H_{\alpha}^{\beta}-H_{\gamma}^{\beta}H_{\beta}^{\alpha} \\
&=H_{\tilde{\gamma}}^{\alpha}H_{\tilde{\gamma}}^{\beta}-
H_{\tilde{\gamma}}^{\alpha}H_{\alpha}^{\beta}-H_{\tilde{\gamma}}^{\beta}H_{\beta}^{\alpha}
\end{aligned}
\end{gather}
where $\alpha, \beta, \gamma, \tilde{\gamma}$ are all distinct.
Then it is a simple check that (\ref{3B}) is satisfied due to
(\ref{2B}) and (\ref{4B}).Thus the associativity conditions for
the algebra (\ref{1B}) are given by equations (\ref{2B}) and
(\ref{4B}). In the particular case when all $C_{\alpha\beta}^{0}$
vanish they are reduced to equations (\ref{associativity3}).

In order to find the coisotropy conditions, let us first compute the Poisson brackets
$\{ f_{\alpha\beta},f_{\gamma\delta} \}$ for pairs of Hamiltonians corresponding to
indices $\alpha,\beta,\gamma,\delta$ all distinct. Since $x_{\alpha},p_{\alpha}$,
for $\alpha=1,2,\ldots,n$ are pairs of conjugate canonical variables, we have
\begin{gather}
\begin{aligned}
\label{5B} &\{ f_{\alpha\beta},f_{\gamma\delta} \} =
\left(\frac{\partial H_{\delta}^{\gamma}} {\partial
x_{\beta}}-\frac{\partial H_{\alpha}^{\beta}} {\partial
x_{\gamma}}\right)f_{\alpha\delta}+\left(\frac{\partial
H_{\delta}^{\gamma}} {\partial x_{\alpha}}-\frac{\partial
H_{\beta}^{\alpha}}
{\partial x_{\gamma}}\right)f_{\beta\delta} \\
&+\left(\frac{\partial H_{\gamma}^{\delta}} {\partial
x_{\beta}}-\frac{\partial H_{\alpha}^{\beta}} {\partial
x_{\delta}}\right)f_{\alpha\gamma}+\left(\frac{\partial
H_{\gamma}^{\delta}} {\partial x_{\alpha}}-\frac{\partial
H_{\beta}^{\alpha}}
{\partial x_{\delta}}\right)f_{\beta\gamma} \\
&+\left(-H_{\alpha}^{\delta}\frac{\partial
H_{\alpha}^{\beta}}{\partial x_{\gamma}}
-H_{\alpha}^{\gamma}\frac{\partial H_{\alpha}^{\beta}}{\partial
x_{\delta}} +H_{\gamma}^{\delta}\frac{\partial
H_{\alpha}^{\beta}}{\partial x_{\gamma}}
-H_{\delta}^{\gamma}\frac{\partial H_{\alpha}^{\beta}}{\partial
x_{\delta}} +H_{\alpha}^{\gamma}\frac{\partial
H_{\gamma}^{\delta}}{\partial x_{\beta}}
+H_{\alpha}^{\delta}\frac{\partial H_{\delta}^{\gamma}}{\partial
x_{\beta}}\right)p_{\alpha} \\
&+\left(-H_{\delta}^{\alpha}\frac{\partial
H_{\alpha}^{\beta}}{\partial x_{\gamma}}
-H_{\delta}^{\beta}\frac{\partial H_{\beta}^{\alpha}}{\partial
x_{\gamma}} +H_{\delta}^{\beta}\frac{\partial
H_{\delta}^{\gamma}}{\partial x_{\alpha}}
+H_{\delta}^{\alpha}\frac{\partial H_{\delta}^{\gamma}}{\partial
x_{\beta}} -H_{\alpha}^{\beta}\frac{\partial
H_{\delta}^{\gamma}}{\partial x_{\alpha}}
-H_{\beta}^{\alpha}\frac{\partial H_{\delta}^{\gamma}}{\partial
x_{\beta}}\right)p_{\delta} \\
&-\left(-H_{\beta}^{\delta}\frac{\partial
H_{\beta}^{\alpha}}{\partial x_{\gamma}}
-H_{\beta}^{\gamma}\frac{\partial H_{\beta}^{\alpha}}{\partial
x_{\delta}} +H_{\gamma}^{\delta}\frac{\partial
H_{\beta}^{\alpha}}{\partial x_{\gamma}}
+H_{\gamma}^{\delta}\frac{\partial H_{\beta}^{\alpha}}{\partial
x_{\delta}} +H_{\beta}^{\gamma}\frac{\partial
H_{\gamma}^{\delta}}{\partial x_{\alpha}}
+H_{\beta}^{\delta}\frac{\partial H_{\delta}^{\gamma}}{\partial
x_{\alpha}}\right)p_{\beta} \\
&+\left(-H_{\gamma}^{\alpha}\frac{\partial
H_{\alpha}^{\beta}}{\partial x_{\delta}}
-H_{\gamma}^{\beta}\frac{\partial H_{\beta}^{\alpha}}{\partial
x_{\delta}} +H_{\gamma}^{\beta}\frac{\partial
H_{\gamma}^{\delta}}{\partial x_{\alpha}}
+H_{\gamma}^{\alpha}\frac{\partial H_{\gamma}^{\delta}}{\partial
x_{\beta}} -H_{\alpha}^{\beta}\frac{\partial
H_{\gamma}^{\delta}}{\partial x_{\alpha}}
-H_{\beta}^{\alpha}\frac{\partial H_{\gamma}^{\delta}}{\partial
x_{\beta}}\right)p_{\gamma} \\
\end{aligned}
\end{gather}
where there is not summation over repeated indices. The Poisson
brackets of Hamiltonians having equal indices , like $\{
f_{\alpha\beta},f_{\beta\delta} \}$, may be obtained from this
formula by means of the following substitutions:
$H_{\beta}^{\beta}=0$ and $ f_{\beta\beta}\rightarrow
p_{2\beta}+f_{11\beta}+v_{\beta}p_{\beta}+u_{\beta}$ where
$f_{11\beta}=p_{\beta}^2-p_{2\beta}-v_{\beta}p_{\beta}-u_{\beta}$
is the lowest Hamiltonian for the $\beta-th$ dKP hierarchy.

The coisotropy condition requires that the r.h.s of (\ref{5B})
vanishes for the values of $(x_{\alpha},p_{\alpha})$ for which
$f_{\alpha\beta}=0$ for all $\alpha\neq{\beta}$. In the case when
all indices in (\ref{5B}) are distinct it is satisfied if the
coefficients in front of
$p_{\alpha},p_{\beta},p_{\gamma},p_{\delta}$ all vanish. It is not
difficult to check that all these equations are satisfied in
virtue of the system
\begin{gather}
\begin{aligned}
\label{6B}
&H_{\alpha}^{\delta}\frac{\partial H_{\alpha}^{\beta}}{\partial x_{\gamma}}
+H_{\alpha}^{\gamma}\frac{\partial H_{\alpha}^{\beta}}{\partial x_{\delta}}
-H_{\gamma}^{\delta}\frac{\partial H_{\alpha}^{\beta}}{\partial x_{\gamma}} \\
&-H_{\delta}^{\gamma}\frac{\partial H_{\alpha}^{\beta}}{\partial
x_{\delta}} -H_{\alpha}^{\gamma}\frac{\partial
H_{\gamma}^{\delta}}{\partial x_{\beta}}
-H_{\alpha}^{\delta}\frac{\partial H_{\gamma}^{\delta}}{\partial x_{\beta}}=0 \\
&                                                                              \\
&\alpha\neq{\beta}\neq{\gamma}\neq{\delta}\neq{\alpha} .
\end{aligned}
\end{gather}
The coisotropy conditions corresponding to pair of Hamiltonians with coinciding
indices , say $ \{ f_{\alpha\beta},f_{\beta\delta} \}$ require that the coefficients
in front of $p_{2\beta}$ ,of $p_{\alpha}$ and $p_{\delta}$, and of $p_{\beta}$
vanish. The first requirement leads to
\begin{equation}
\label{7B} \frac{\partial H_{\beta}^{\delta}}{\partial
x_{\alpha}}-\frac{\partial H_{\beta}^{\alpha}}{\partial
x_{\delta}}=0 \qquad \alpha\neq{\beta}\neq{\delta}\neq{\alpha} .
\end{equation}
The second requirement gives equation (\ref{6B}) with $\beta\neq{\gamma}$, and
finally the  third requirement gives the equation relating $H_{\beta}^{\alpha}$
with the functions $v_{\beta},u_{\beta}$ for the $\beta-th$ KP hierarchy.

So the coisotropy conditions contain the subset of equations (\ref{6B}),(\ref{7B})
containing only the functions $H_{\beta}^{\alpha}$ for $\alpha\neq{\beta}$. Using
(\ref{7B}), one shows that equations (\ref{6B}) can be recasted in the form
\begin{equation}
\label{8B}
\frac{\partial}{\partial x_{\beta}}\left(H_{\alpha}^{\gamma}H_{\alpha}^{\delta}
-H_{\alpha}^{\delta}H_{\delta}^{\gamma}-H_{\alpha}^{\gamma}H_{\gamma}^{\delta}
\right) =0 .
\end{equation}
Thus this subset of coisotropy conditions is equivalent to the
associativity conditions (\ref{associativity3}) and to the
exactness conditions (\ref{7B}).

Note that in terms of the functions $F_{\alpha}$ introduced in Sec.4 one has
\begin{gather}
\begin{aligned}
\label{9B}
\{ f_{\alpha\beta},f_{\gamma\delta} \} &= \frac{{\partial}^2 (F_{\delta}-F_{\alpha})}
{\partial x_{\beta}\partial x_{\gamma}}f_{\alpha\delta}
+\frac{{\partial}^2 (F_{\delta}-F_{\beta})}
{\partial x_{\alpha}\partial x_{\beta}}f_{\beta\delta}  \\
&+\frac{{\partial}^2 (F_{\gamma}-F_{\alpha})} {\partial
x_{\beta}\partial x_{\delta}}f_{\alpha\gamma} +\frac{{\partial}^2
(F_{\gamma}-F_{\beta})} {\partial x_{\alpha}\partial
x_{\delta}}f_{\beta\gamma}  .
\end{aligned}
\end{gather}
This representation of the dDarboux system is the quasiclassical
limit of the operator representation of the original Darboux
system.

The coisotropic submanifold $\Gamma$ defined by the equations
\begin{equation}
\label{10B} f_{\alpha\beta}=
p_{\alpha}p_{\beta}-H_{\alpha}^{\beta}p_{\alpha}
-H_{\beta}^{\alpha}p_{\beta}-C_{\alpha\beta}^{0}=0  \qquad
\alpha\neq{\beta}
\end{equation}
and the quadrics defined by the associativity conditions have
quite remarkable properties. First, the quadrics (\ref{10B}) are
transformed into quadrics of the same type under a Cremona
transformation $p_{\alpha}\rightarrow
\xi_{\alpha}=\frac{1}{p_{\alpha}}$. Indeed one gets
\begin{equation*}
\xi_{\alpha} \xi_{\beta}-\widetilde{H}_{\alpha}^{\beta}
\xi_{\alpha}- \widetilde{H}_{\beta}^{\alpha}
\xi_{\beta}-\widetilde{C}_{\alpha \beta}^{0}=0
\end{equation*}
where
\begin{equation*}
\widetilde{H}_{\alpha}^{\beta}=
-\frac{H_{\alpha}^{\beta}}{C_{\alpha \beta}^{0}} \qquad
\widetilde{C}_{\alpha \beta}^{0}= \frac{1}{C_{\alpha \beta}^{0}}.
\end{equation*}
In the particular case $C_{\alpha\beta}^{0}=0$ the Cremona's transformation linearises
the quadric (\ref{10B}), since in this case
\begin{equation}
\label{12B}
H_{\alpha}^{\beta}\xi_{\beta}+H_{\beta}^{\alpha}\xi_{\alpha}-1=0
\qquad \textup{for} \qquad \alpha,\beta=1,2,\ldots,n  \quad
\alpha\neq{\beta} .
\end{equation}
So for the dDarboux system any section of the coisotropic
submanifold with $x_{\alpha}=const$ is , in fact, the intersection
of the planes (\ref{12B}).

Secondly, the equations defining the associativity quadrics can be rewritten as
\begin{equation*}
\frac{H_{\gamma}^{\beta}}{H_{\alpha}^{\beta}}+
\frac{H_{\beta}^{\gamma}}{H_{\alpha}^{\gamma}} - 1 =0  \qquad
\alpha\neq{\beta}\neq{\gamma}\neq{\alpha} .
\end{equation*}
In the case $n=3$ , in terms of the variables $z_{1},z_{2},z_{3}$ defined by
\begin{equation*}
z_{1}=\frac{H_{2}^{1}}{H_{3}^{1}} \qquad  z_{2}=\frac{H_{3}^{2}}{H_{2}^{1}} \qquad
z_{3}=\frac{H_{1}^{3}}{H_{2}^{3}}
\end{equation*}
the above equations become
\begin{equation*}
z_{1}+\frac{1}{z_{2}}=1  \qquad   z_{2}+\frac{1}{z_{3}}=1  \qquad
z_{3}+\frac{1}{z_{1}}=1
\end{equation*}
with the obvious constraint
\begin{equation*}
 z_{1}z_{2}z_{3}=-1 .
\end{equation*}
These equations define a curve in $R^{3}$ which is the
intersection of three cylinders which are generated by the above
hyperbolas.\\

{\bf Acknowledgements.} The authors are grateful to A. Moro for
the help in preparation of the paper. The work has been partly
supported by the grants COFIN 2004 "Sintesi", and COFIN 2004
"Nonlinear Waves and Integrable systems"

\end{document}